\documentclass{aa}

\usepackage[varg]{txfonts}
\usepackage{graphicx}
\usepackage{natbib}
\usepackage{xcolor} % For \textcolor command
\usepackage{soul}
\usepackage[hyperindex,breaklinks=true,colorlinks,linkcolor=blue,citecolor=blue,urlcolor=blue]{hyperref}

\graphicspath{{Images/}}

\newcommand\ve[1]{\textbf{#1}}

\providecommand{\sorthelp}[1]{}

\title{Statistical description of dust polarized emission from the diffuse interstellar medium}
\subtitle{A RWST approach}

\author{B. Regaldo-Saint Blancard\inst{\ref{lra},\ref{obs}}
\and F. Levrier\inst{\ref{lra}}
\and E. Allys\inst{\ref{lra}}
\and E. Bellomi\inst{\ref{lra},\ref{obs}}
\and F. Boulanger\inst{\ref{lra}}}

\institute{{Laboratoire de Physique de l'École Normale Supérieure, ENS, Université PSL, CNRS, Sorbonne Université, Université de Paris, F-75005 Paris, France\\ \email{bruno.regaldo@phys.ens.fr}\label{lra}} \and {Observatoire de Paris, PSL University, Sorbonne Université, LERMA, 75014 Paris, France\label{obs}}}

\date{Received: March 27, 2020 /
Accepted: August 27, 2020}

\abstract {The statistical characterization of the diffuse magnetized interstellar medium (ISM) and Galactic foregrounds to the cosmic microwave background (CMB) poses a major challenge. To account for their non-Gaussian statistics, we need a data analysis approach capable of efficiently quantifying statistical couplings across scales. This information is encoded in the data, but most of it is lost when using conventional tools, such as one-point statistics and power spectra. The wavelet scattering transform (WST), a low-variance statistical descriptor of non-Gaussian processes introduced in data science, opens a path towards this goal. To establish the methodology, we applied the WST to noise-free maps of dust polarized thermal emission computed from a numerical simulation of magnetohydrodynamical (MHD) turbulence in the diffuse ISM. We analyzed normalized complex Stokes maps and maps of the polarization fraction and polarization angle. The WST yields a few thousand coefficients; some of them measure the amplitude of the signal at a given scale, and the others characterize the couplings between scales and orientations. The dependence on orientation can be fitted with the reduced wavelet scattering transform (RWST), an angular model introduced in previous works for total intensity maps. The RWST provides a statistical description of the polarization maps, quantifying their multiscale properties in terms of isotropic and anisotropic contributions. It allowed us to exhibit the dependence of the map structure on the orientation of the mean magnetic field and to quantify the non-Gaussianity of the data. We also used RWST coefficients, complemented by additional constraints, to generate random synthetic maps with similar statistics. Their agreement with the original maps demonstrates the comprehensiveness of the statistical description provided by the RWST. This work is a step forward in the analysis of observational data and the modeling of CMB foregrounds. We also release {\tt PyWST}, a public Python package to perform WST and RWST analyses of two-dimensional data at: \url{https://github.com/bregaldo/pywst}.}
\bibpunct{(}{)}{;}{a}{}{,} % to follow the A&A style

\keywords{ISM: dust, extinction - ISM: magnetic fields - turbulence - methods: statistical - polarization}

\begin{document}

\maketitle

%%%%%%%%%%%%%%%%%%%%%%%%%%%%%%%%%%%%%%%%%%%%%%%%%%%%
%%%%%%%%%%%%%%%%%%%%%%%%%%%%%%%%%%%%%%%%%%%%%%%%%%%%
%% SECTION 0: NOTES
%%%%%%%%%%%%%%%%%%%%%%%%%%%%%%%%%%%%%%%%%%%%%%%%%%%%
%%%%%%%%%%%%%%%%%%%%%%%%%%%%%%%%%%%%%%%%%%%%%%%%%%%%

%\setcounter{section}{-1}

%%%%%%%%%%%%%%%%%%%%%%%%%%%%%%%%%%%%%%%%%%%%%%%%%%%%
%%%%%%%%%%%%%%%%%%%%%%%%%%%%%%%%%%%%%%%%%%%%%%%%%%%%
%% SECTION 1: INTRODUCTION
%%%%%%%%%%%%%%%%%%%%%%%%%%%%%%%%%%%%%%%%%%%%%%%%%%%%
%%%%%%%%%%%%%%%%%%%%%%%%%%%%%%%%%%%%%%%%%%%%%%%%%%%%

\section{Introduction}

The interstellar medium (ISM) is a beautifully complex physical system, in which gas particles and dust grains, coupled to a pervasive magnetic field, experience turbulent motions across a vast range of scales~\citep{Draine2011,HennebelleFalgarone2012}. Nonspherical grains tend to align locally with their longest axis perpendicular to the Galactic magnetic field~\citep{Andersson2015,Reissl20}, leading to polarization in extinction in the visible and in emission in the submillimeter~\citep{DraineFraisse2009,Guillet2018}. The multiscale filamentary structure of diffuse interstellar matter is spectacularly illustrated by {\it Herschel} observations of dust emission~\citep{Miville10}. Observations of dust polarization provide an additional perspective because they probe the orientation of magnetic fields. {\it Planck} has provided us with the first all-sky survey of polarized dust emission, opening the path to statistical studies~\citep{planck2016-l11B}. This broad view is being complemented by observations at higher angular resolution, which are carried out by the balloon-borne experiments  BLASTPOL~\citep{Fissel2016}  and  PILOT~\citep{Mangilli19},  the  far-IR  HAWC+  camera  onboard SOFIA~\citep{Chuss19} and imaging at sub-mm/mm wavelengths from large single-dish telescopes~\citep{Ritacco20} and ALMA~\citep{Hull17}. These observations all contribute to a common scientific goal: understanding the role turbulence and magnetic fields play along the star formation process, from the diffuse interstellar medium to molecular clouds and protostellar cores. In this ambitious endeavor, all projects face the difficulty of inferring information on magnetic fields from observations of Stokes parameters. 

Over the last few years, studies of the magnetized and turbulent ISM have become closely entwined with a major goal of observational cosmology that is the detection of the primordial gravitational wave (GW) signal from the early Universe's inflationary era~\citep{guth1981,linde1982}. The reason for this entanglement lies in the superposition  between the expected signal from these GWs on the one hand, imprinted in the curl-like "B-mode" part of the polarization of the cosmic microwave background~\citep[CMB,][]{kamionkowski1997}, and the polarized emission from the Galaxy on the other hand ~\citep{pb2015,planck2014-XXX,planck2016-l11A}. The current constraints on the cosmological B-mode tensor-to-scalar ratio $r<0.06$~\citep[95\% confidence level,][]{Bicep2018limit} are expected to be significantly improved by the next generation of CMB experiments, such as ACT~\citep{naess2014}, SPIDER~\citep{fraisse2013}, LiteBIRD~\citep{ishino2016}, PICO~\citep{PICO-2018}, the Simons Observatory~\citep{Simons-Observatory-2019}, and CMB-S4~\citep{CMB-S4-Science-Case}, with a detection limit goal of $r\simeq 10^{-3}$. However, any claim to the detection of the cosmological B-mode signal will have to be critically assessed against alternative explanations involving Galactic foregrounds.

An overarching challenge for studying both interstellar turbulence and CMB foregrounds is the need for a new approach to data analysis, which is able to efficiently capture statistical couplings across scales. This information is encoded in the data, but most of it is lost when one resorts to such classical tools as one-point statistics and power spectra. 

To characterize magnetized interstellar turbulence, as well as to produce realistic synthetic dust polarization sky maps for CMB data analysis, we aim at a statistical description of the polarized Galactic emission maps that encompasses non-Gaussian characteristics. Recent advances in data science open up a new path towards this goal. \cite{Allys2019} provided the first astrophysical application of the wavelet scattering transform (WST), a low-variance statistical description of non-Gaussian processes~\citep{Mallat2012} inspired by convolutional neural networks but that does not require any training stage~\citep{Bruna2013}. They applied the WST to column density maps inferred from magnetohydrodynamical (MHD) simulations and to an {\it Herschel} observation of the thermal emission from Galactic dust.  The physical regularity of the maps led them to introduce the reduced wavelet scattering transform (RWST), a statistical description of reduced dimensionality, obtained through a fit of the angular dependencies of the WST coefficients.

In this paper, we extend the WST and RWST analyses to maps of linearly polarized emission, which is represented by Stokes $Q$ and $U$ maps\footnote{The last Stokes parameter $V$ measuring the intensity of circularly polarized light is ignored in this paper as it is expected to be negligible in the diffuse interstellar medium at frequencies of a few hundred GHz (\citet{Siebenmorgen2014}).}. We begin by presenting the properties of these Stokes maps and define three polarization variables of interest built from the full $(I,Q,U)$ maps. We then recall the definition of the WST for real maps and extend it to complex maps. In order to focus on the properties of this new statistical description of polarization data, we work with data sets built from a noise-free MHD simulation, from which we compute simulated Stokes maps. We identify regular patterns in the angular dependencies of the WST coefficients for these maps, leading us to define a RWST model for polarization maps. We give interpretations of the RWST coefficients that highlight the impact of the orientation of the mean magnetic field on the statistical properties of polarization maps. We also show how these coefficients quantitatively exhibit the non-Gaussianity of these maps. We finally assess the exhaustiveness of RWST descriptions of polarization maps by generating random maps from these statistical coefficients.

The paper is organized as follows: In Sect.~\ref{section2}, we recall the basic properties of Stokes maps of polarized thermal emission from dust and introduce the set of polarization variables on which we define the WST. In Sect.~\ref{section3}, we present the MHD simulation and the polarization maps derived from it, exhibit regularities in the WST coefficients for these maps, and define the RWST model. In Sect.~\ref{section4}, we give interpretations for a subset of the RWST coefficients. We present synthetic realizations of random polarization maps based on the RWST coefficients in Sect.~\ref{SectionSyntheses}, and summarize our conclusions in Sect.~\ref{SectionConclusions}. We also provide a public Python package to perform WST and RWST analyses of two-dimensional data called {\tt PyWST} at: \url{https://github.com/bregaldo/pywst}.

%%%%%%%%%%%%%%%%%%%%%%%%%%%%%%%%%%%%%%%%%%%%%%%%%%%%
%%%%%%%%%%%%%%%%%%%%%%%%%%%%%%%%%%%%%%%%%%%%%%%%%%%%
%% SECTION 2: 
%%%%%%%%%%%%%%%%%%%%%%%%%%%%%%%%%%%%%%%%%%%%%%%%%%%%
%%%%%%%%%%%%%%%%%%%%%%%%%%%%%%%%%%%%%%%%%%%%%%%%%%%%

\section{Statistical description of polarization maps with the WST}
\label{section2}

We first discuss the properties of Stokes $I$, $Q$, $U$ maps of the polarized emission of dust before defining convenient transformations of these observables. We then explain how to apply the WST to these transformed maps.

\subsection{Properties of $I$, $Q$, $U$  maps}
\label{PropertiesOfStokesMaps}

In the diffuse interstellar medium and at the frequencies of a few hundred GHz, we observe the thermal emission of large grains of dust ~\citep[typically grains of radii greater than 50 nm, see e.g.,][]{Draine2011}. These large grains are in thermal equilibrium, absorbing the light of surrounding stars and emitting a black body radiation at typical temperatures of a few tens of K. This emission is polarized due to the statistical alignment between the orientation of aspherical dust grains and the local magnetic field~\citep{Andersson2015}, leading to preferential directions of absorption and emission of light, and thus to a polarization of the signal~\citep{planck2014-XIX}.

\begin{figure}[h!]
	\centering
	\includegraphics[width=\hsize]{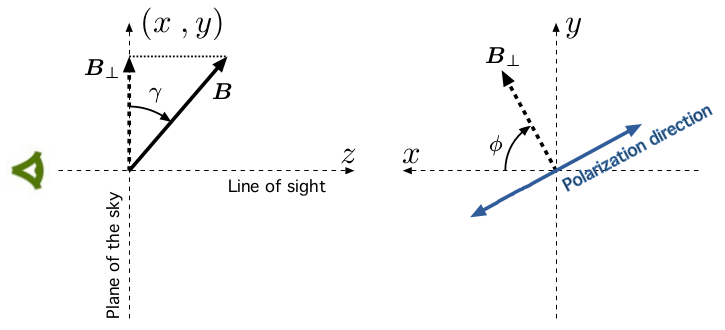}
	\caption{Definition of the angles $\gamma$ and $\phi$ related to the orientation of the local magnetic field with respect to the plane of the sky when the $z$-axis corresponds to the line of sight~\citep[adapted from][]{planck2014-XX}. The line of sight points away from the observer and $\phi$ is counted positively clockwise from the $x$ direction in the {\tt HEALPix} convention.}
	\label{BAngles}
\end{figure}

Taking these processes into account, a physical model of radiative transfer has been built~\citep[see derivation and references in][]{planck2014-XX}. This model provides analytical expressions at a given submillimeter frequency  $\nu$ for the Stokes parameters $I,Q,$ and $U$:
\begin{align}
	I &= \int S_\nu e^{-\tau_\nu}\left[1-p_0\left(\cos^2\gamma - \frac{2}{3}\right)\right]\mathrm{d}\tau_\nu,\label{eqIRaw}\\
	Q &= \int p_0S_\nu e^{-\tau_\nu}\cos\left(2\phi\right)\cos^2\gamma\mathrm{d}\tau_\nu,\label{eqQRaw}\\
	U &= \int p_0S_\nu e^{-\tau_\nu}\sin\left(2\phi\right)\cos^2\gamma\mathrm{d}\tau_\nu,\label{eqURaw}
\end{align}
where $S_\nu$ is the source function which is assumed here to be that of a black body so that $S_\nu = B_\nu(T_d)$ with $T_d$ the dust temperature, $\tau_\nu$ is the optical depth, $p_0$ is an intrinsic polarization fraction parameter, $\gamma$ is the angle that the local magnetic field makes with the plane of the sky, and $\phi$ is the angle that the projection of the local magnetic field on the plane of the sky makes with the $x$ direction when $z$ is the line of sight in the {\tt HEALPix}\footnote{\tt https://healpix.jpl.nasa.gov.}~\citep{gorski2005} convention (see Fig. \ref{BAngles}).

In the following, we assume uniform values for the intrinsic polarization $p_0$ and the dust temperature $T_d$, an optically thin medium, and an infinitesimal optical depth equal to ${\mathrm{d}\tau_\nu=\sigma_\nu n_\mathrm{H}\mathrm{d}z}$ where $\sigma_\nu$ is the dust opacity (assumed uniform) and $n_\mathrm{H}$ is the total gas density. In this case, Eqs.~(\ref{eqIRaw})-(\ref{eqURaw}) become:
\begin{align}
	I &=  S_\nu \sigma_\nu\left[N_\mathrm{H} - p_0\int n_\mathrm{H}\left(\cos^2\gamma-\frac{2}{3}\right)\mathrm{d}z\right],\label{eqI}\\
	Q &= p_0\sigma_\nu S_\nu \int n_\mathrm{H}\cos\left(2\phi\right)\cos^2\gamma\mathrm{d}z \label{eqQ},\\
	U &= p_0\sigma_\nu S_\nu \int n_\mathrm{H}\sin\left(2\phi\right)\cos^2\gamma\mathrm{d}z. \label{eqU}
\end{align}

From these formulæ, Stokes parameters $I$, $Q$, and $U$ appear to be proportional to an integration along the line of sight of quantities depending on the same three variables: the total gas density $n_\mathrm{H}$, and the angles $\gamma$ and $\phi$. Therefore we expect a priori some statistical dependencies between $I$, $Q$, and $U$ maps, and a statistical description of these Stokes maps should take into account these dependencies.

These expressions also underline how closely related Stokes $Q$ and $U$ are. Indeed they are defined with respect to a given reference direction~\cite[see for example][]{LandiDeglinnocenti2004}, and any rotation of this reference direction by a given angle $\varphi$ transforms $Q$ and $U$ into rotated quantities $Q^\prime$ and $U^\prime$ as follows\footnote{These relations hold for an angle $\varphi$ that is counted positively clockwise from the $x$ direction when the $z$-axis is the line of sight in the {\tt HEALPix} convention. See Fig. \ref{BAngles}.}:
\begin{align}
    Q^\prime &= \cos(2\varphi)Q + \sin(2\varphi)U,\\
    U^\prime &= -\sin(2\varphi)Q + \cos(2\varphi)U.
\end{align}
These relations can be written in a more compact form introducing complex quantities:
\begin{equation}
    \label{QUtransfo}
	Q^\prime + iU^\prime = \left(Q+iU\right)e^{-2i\varphi}.
\end{equation}
As a consequence, the global phase of $Q+iU$ complex maps is directly related to the reference frame in which we define the angular variables measuring the orientation of the local magnetic field within the plane of the sky. Because of the previous transformation properties, the complex variable $Q+iU$ is more apt to represent linear polarization than $Q$ and $U$ separately~\citep[see e.g.,][]{2004A&A...427..549H}. Moreover it will be useful to define a statistical description of polarization maps that is independent from this choice of reference frame as we discuss in Sect.~\ref{wstSubsection}.

Finally, let us discuss the non-Gaussianity of Stokes maps $I$, $Q$, and $U$. We know that the gas and the magnetic field in the diffuse interstellar medium are highly turbulent~\citep{HennebelleFalgarone2012}, and the nonlinearity involved in the MHD equations that describe their dynamics couples scales. Therefore we expect the statistical properties of the sky maps of $I$, $Q$, and $U$ to be highly non-Gaussian. Unless one can find a mapping of the data that leads to Gaussian statistics, a mere statistical description in terms of power spectra measurements cannot be sufficient to statistically characterize these maps.

\subsection{Transformation of Stokes maps}
\label{transfoStokes}

Stokes maps $I$ and $Q+iU$ are raw observables of the polarized emission of dust that we first want to transform. Our goal is twofold: to simplify the statistical properties of these observables and to ease the physical interpretation of their properties. In this work, we consider three dimensionless variables based on Stokes maps $I$ and $Q+iU$ to build and interpret a statistical description of polarized light: normalized Stokes variable $\tilde{Q}+i\tilde{U}$, polarization fraction $p$ and complex polarization angle $\exp(2i\psi)$.

We first define the normalized Stokes complex variable ${\tilde{Q}+i\tilde{U}}$ as follows:
\begin{align}
    \tilde{Q} + i\tilde{U} = \frac{Q+iU}{I + P},
\end{align}
with $P = |Q+iU|$ the polarized intensity. Such a definition roughly disentangles the structure of the magnetic field and the structure of dust density at the lowest order. Indeed, assuming a constant orientation of the magnetic field along the line of sight, one gets using Eqs.~(\ref{eqI}), (\ref{eqQ}), and (\ref{eqU}):
\begin{align}
    \tilde{Q}+i\tilde{U}=\frac{3p_0}{3+2p_0}\cos^2\gamma\exp(2i\phi),
\end{align}
which is independent of the density field $n_\mathrm{H}$.

The polarization fraction $p$ and the polarization angle $\psi$ follow the usual definitions:
\begin{align}
    p &= \frac{|Q+iU|}{I},
\end{align}
and
\begin{align}
    \psi &= \frac1{2}\mathrm{atan2} (U, Q),
\end{align}
where $\mathrm{atan2}(b, a)$ is the function that returns the angle restricted to $(-\pi,\pi]$ between the positive $x$-axis and the ray to the point $(a,b)$ in the Euclidean plane. Because of the nonlinearity of the $\mathrm{atan2}$ function, we expect the statistics of $\psi$ maps to be unnecessarily complicated compared to those of the original Stokes maps. We thus choose to analyze maps of the complex variable $\exp(2i\psi)$. This variable also avoids dealing with discontinuities of $\psi$ maps that appear where the polarization angle approaches $\pm\pi/2$. Moreover, the statistical properties of $\exp(2i\psi)$ are easier to compare to those of $\tilde{Q} + i\tilde{U}$ since $\tilde{Q}+i\tilde{U}\approx p\exp(2i\psi)$ when $P \ll I$ (which is typically the case).

\subsection{WST statistical description}
\label{wstSubsection}

The WST is a tool from data science that computes statistical descriptions of images that can efficiently discriminate between images which have identical power spectra but distinct higher order moments~\citep{Bruna2013}. An advantage of this description compared to the direct estimation of higher order moments is that the estimators of the WST coefficients have a low variance and thus do not need a large amount of samples for an accurate estimation. This is because the WST relies on nonexpansive operators \citep[see][]{Bruna2013}. Originally designed to understand the properties of deep convolutional neural networks which have proved successful for classification problems~\citep{LeCun2010,Krizhevsky2012}, this tool also provides a convenient statistical description of astronomical observations of the interstellar medium for total intensity maps~\citep{Allys2019}. In this subsection we use the WST to describe the statistical properties of the polarization related maps discussed in the previous section: complex maps $\tilde{Q} + i\tilde{U}$ and $\exp(2i\psi)$, and real map $p$.

Let us briefly recall the definition of the WST of a given 2D real field $I(\ve{x})$ that is statistically homogeneous\footnote{In this work, we restrict our analysis to statistically homogeneous data for simplicity, but this is not a limitation of the methodology as the WST can also be computed locally~\citep[see e.g.,][]{Allys2019}.}. For a more complete presentation we refer to \citet{Allys2019} and \citet{Bruna2013}. The WST relies on convolutions of the target field $I(\ve{x})$ with a set of $J\times \Theta$ Morlet wavelets $\psi_{j,\theta}$ with ${0\leq j \leq J - 1}$ and $\theta\in\{k\pi/\Theta, 0\leq k \leq \Theta - 1\}$, with $J$ and $\Theta$ two integers. These wavelets $\psi_{j, \theta}$ are the result of the dilation by a factor $2^j$ and the rotation by an angle $\theta$ of a mother Morlet wavelet $\Psi$:
\begin{equation}
	\psi_{j, \theta} = 2^{-2j}\Psi(2^{-j}r_{\theta}^{-1}\ve{x}).
\end{equation}
A Morlet wavelet is essentially a plane wave modulated by a Gaussian envelope. It is a quite general kind of wavelet to study physical fields and it provides a good angular selectivity~\citep{Farge1992}. We recall the definition of the mother Morlet wavelet $\Psi$:
\begin{equation}
\label{eq_motherwavelet}
	\Psi\left(\ve{x}\right) = \alpha\left(e^{i\ve{k}_0\cdot\ve{x}} - \beta\right)e^{-|\ve{x}|^2/2\sigma^2},
\end{equation}
with $\alpha$ and $\beta$ two constants that are adjusted to ensure a zero mean and a unit $L^1$ norm, $\ve{k}_0 = k_0\ve{e}_x$ the wave vector of the plane wave factor, and $\sigma$ the standard deviation of the Gaussian envelope\footnote{In practice we choose $k_0=3\pi/4\,\mathrm{pixel}^{-1}$ and $\sigma = 0.8\,\mathrm{pixel}$ as in \citet{Bruna2013}.}$^{,}$\footnote{Actually, the Gaussian envelope has an elliptical shape that is not included in Eq.~(\ref{eq_motherwavelet}) for simplicity. This elliptical shape enhances the angular selectivity of the wavelets.}.
Convolving a field $I(\ve{x})$ with a wavelet $\psi_{j, \theta}$ corresponds to a local band-pass filtering of the field at frequencies centered on a mode $\ve{k}_{j,\theta} = k_02^{-j}\left(\cos\left(\theta\right)\hat{\ve{e}}_x + \sin\left(\theta\right)\hat{\ve{e}}_y\right)$. $J$ and $\Theta$ constants thus correspond respectively to the number of scales and angles that discretize the 2D Fourier space.
For a statistically homogeneous real field $I(\ve{x})$, normalized WST coefficients are defined as follows:
\begin{align}
	&\bar{S}_0 = \langle I \rangle,\label{S0eqReal}\\
	&\bar{S}_1(j_1,\theta_1) = \frac{\langle|I\star\psi_{j_1, \theta_1}|\rangle}{\langle I \rangle},\label{S1eqReal}\\ &\text{ with } 0 \leq j_1 \leq J - 1 \text{ and } 0 \leq \theta_1 < \pi,\nonumber\\
	&\bar{S}_2 (j_1,\theta_1,j_2,\theta_2) =\frac{\langle ||I \star \psi_{j_1, \theta_1}|\star\psi_{j_2, \theta_2}|\rangle}{\langle |I\star\psi_{j_1, \theta_1}|\rangle}\label{S2eqReal}, \\ &\text{ with } 0 \leq j_1,j_2 \leq J - 1 \text{ such that } j_2 > j_1 \text{ and } 0 \leq \theta_1,\theta_2 < \pi, \nonumber
\end{align}
where the averaging operator $\langle \cdot \rangle$ here is simply an average over a map, meaning $\langle I \rangle = \frac1{A}\int I(\ve{x})\mathrm{d}^2\ve{x}$ with $A$ the area of integration, and where the $\star$ symbol stands for the convolution operator. The number of WST coefficients depends on the $J$ and $\Theta$ parameters: we have a single $\bar{S}_0$ coefficient, $J\times\Theta$ $\bar{S}_1$ coefficients, and $\Theta^2\times J\times(J-1)/2$ $\bar{S}_2$ coefficients.

These coefficients can be interpreted in the following manner: $\bar{S}_0$ is simply the mean of the field, $\bar{S}_1(j_1,\theta_1)$ is a measure of the amplitude of oscillation of the normalized $I/\langle I\rangle$ field for modes centered on $\ve{k}_{j_1,\theta_1}$, and finally $\bar{S}_2(j_1, \theta_1, j_2, \theta_2)$ characterizes how the normalized amplitude of oscillation of the field for a mode $\ve{k}_{j_1,\theta_1}$ is modulated by a mode of oscillation $\ve{k}_{j_2,\theta_2}$. Accordingly, the $\bar{S}_1(j_1,\theta_1)$ coefficient characterizes the amplitude at a single oriented scale $(j_1,\theta_1)$, while the $\bar{S}_2(j_1, \theta_1, j_2, \theta_2)$ coefficient measures the coupling between two oriented scales $(j_1, \theta_1)$ and $(j_2, \theta_2)$.

We note that even if the $\bar{S}_1$ coefficients characterize the Fourier amplitude of the field under study in the spectral band of the wavelets, they differ in practice from the power spectrum. While the power spectrum can be computed from $L^2$ norms of the wavelet transform, the $\bar{S}_1$ coefficients are computed with $L^1$ norms. One can however recover the power spectrum at a given scale from a quadratic sum of $\bar{S}_1$ and $\bar{S}_2$ coefficients\footnote{See Eq.~(9) in~\citet{Allys2019}.}. Notably, we expect a non-Gaussian field to have higher $\bar{S}_2$ coefficients compared to those of a Gaussian field with identical power spectrum, and this should be counterbalanced by smaller $\bar{S}_1$ coefficients. This is related to the sparsity of the wavelet representation of the data (see \citet{Allys2019} and \citet{Bruna2013} for further discussions).

We can extend the previous definition of normalized WST coefficients to statistically homogeneous complex fields such as $\tilde{Q}+i\tilde{U}$ or $\exp(2i\psi)$ as follows:
\begin{align}
	&\bar{S}_0 = \langle |\tilde{Q}+i\tilde{U}|\rangle,\label{S0eqCplx}\\
	&\bar{S}_1(j_1,\theta_1) = \frac{\langle |(\tilde{Q}+i\tilde{U})\star\psi_{j_1, \theta_1}|\rangle}{\langle |\tilde{Q}+i\tilde{U}| \rangle},\label{S1eqCplx}\\
	&\bar{S}_2 (j_1,\theta_1,j_2,\theta_2) =\frac{\langle ||(\tilde{Q}+i\tilde{U}) \star \psi_{j_1, \theta_1}|\star\psi_{j_2, \theta_2}|\rangle }{\langle |(\tilde{Q}+i\tilde{U})\star\psi_{j_1, \theta_1}|\rangle},\label{S2eqCplx}
\end{align}
where we still have $0\leq j_1,j_2 \leq J - 1$ with $j_2 > j_1$ for $\bar{S}_2$ coefficients, but here the main difference is that $0\leq \theta_1 < 2\pi$ and $0\leq \theta_2 < \pi$. This new range of angles for $\theta_1$ is directly related to the complex nature of the $\tilde{Q}+i\tilde{U}$ variable. For a real signal $I(\ve{x})$, one can easily show that $|I\star\psi_{j_1, \theta_1 + \pi}|=|I\star\psi_{j_1, \theta_1}|$, which explains the range of $\theta_1$ in Eq.~(\ref{S1eqReal}). For a complex signal the previous relation does not hold anymore, and we have to consider rotations of the mother wavelet $\Psi$ for angles between $0$ and $2\pi$~\citep[see][for more details]{Mallat2012}. We note that the range of angles for $\theta_2$ is unchanged because $|(\tilde{Q}+i\tilde{U})\star\psi_{j_1, \theta_1}|$ is a real signal. Thus, for a complex field, we end up with twice as many $\bar{S}_1$ and $\bar{S}_2$ coefficients as for the WST of a real field.

Let us note that this definition of the WST on complex polarization maps $\tilde{Q}+i\tilde{U}$ does not depend on the global phase of $\tilde{Q}+i\tilde{U}$ maps, meaning that the WST coefficients associated to $\tilde{Q}+i\tilde{U}$ maps do not depend on the reference frame mentioned in Sect.~\ref{PropertiesOfStokesMaps}.

In the following, the statistical properties of $p$ maps will be described with the WST coefficients defined in Eqs.~(\ref{S0eqReal}), (\ref{S1eqReal}), and (\ref{S2eqReal}), while the statistical properties of $\tilde{Q} + i\tilde{U}$ and $\exp(2i\psi)$ maps will be described with the WST coefficients defined in Eqs.~(\ref{S0eqCplx}), (\ref{S1eqCplx}), and (\ref{S2eqCplx}).

%%%%%%%%%%%%%%%%%%%%%%%%%%%%%%%%%%%%%%%%%%%%%%%%%%%%
%%%%%%%%%%%%%%%%%%%%%%%%%%%%%%%%%%%%%%%%%%%%%%%%%%%%
%% SECTION 3: 
%%%%%%%%%%%%%%%%%%%%%%%%%%%%%%%%%%%%%%%%%%%%%%%%%%%%
%%%%%%%%%%%%%%%%%%%%%%%%%%%%%%%%%%%%%%%%%%%%%%%%%%%%

\section{Simplification of the WST statistical descriptions for simulated Stokes maps with the RWST}
\label{section3}

\begin{figure*}
	\centering
	\includegraphics[width=\textwidth]{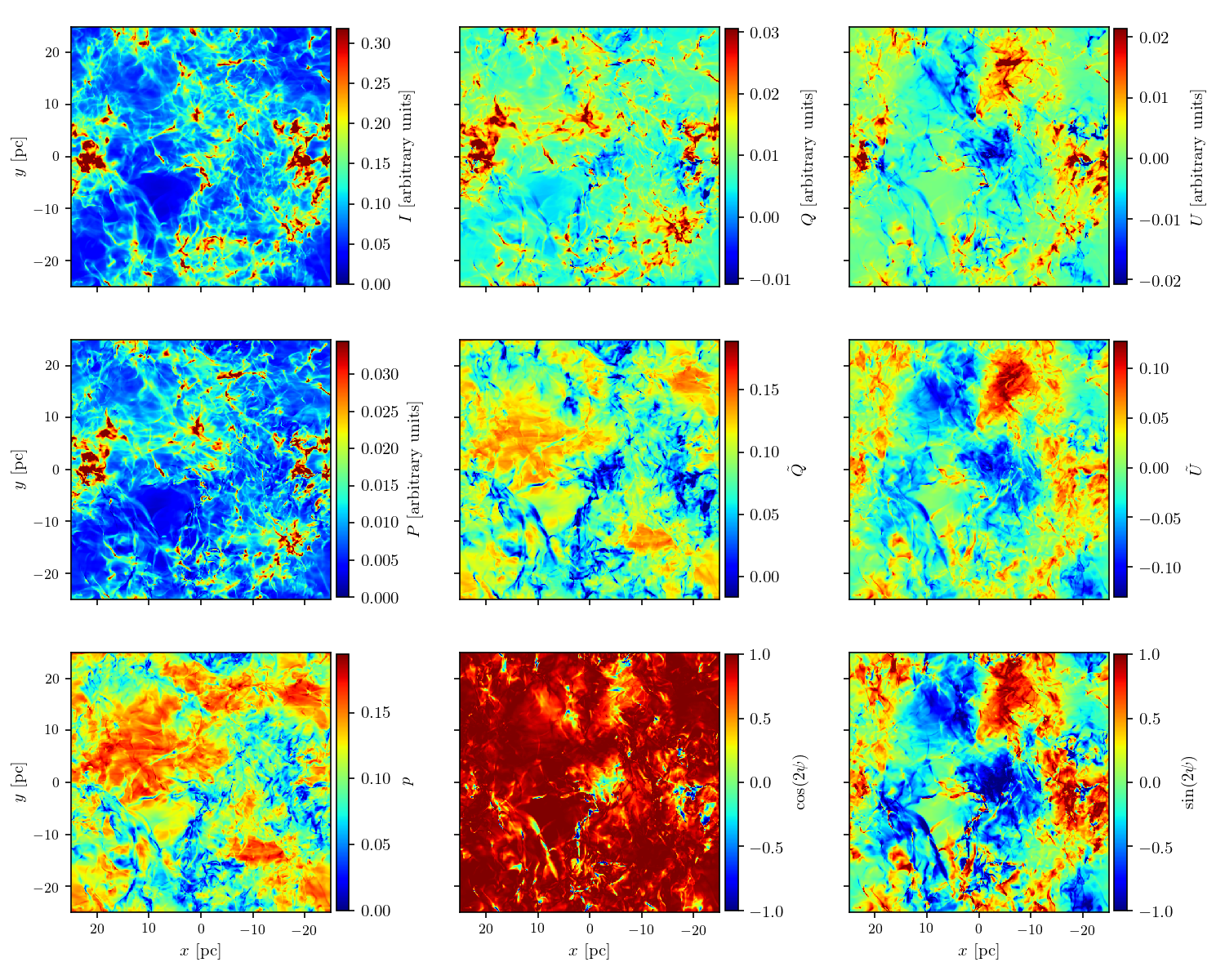}
	\caption{Examples of $I_{\bot}$, $Q_{\bot}$, $U_{\bot}$, $P_{\bot}$, $\tilde{Q}_{\bot}$, $\tilde{U}_{\bot}$, $p_{\bot}$, $\cos(2\psi)_{\bot}$, and $\sin(2\psi)_{\bot}$ maps (from top to bottom and left to right) that are built from a given snapshot of the MHD simulation.}
	\label{datasetOrtho}
\end{figure*}

\begin{figure*}
	\centering
	\includegraphics[width=\textwidth]{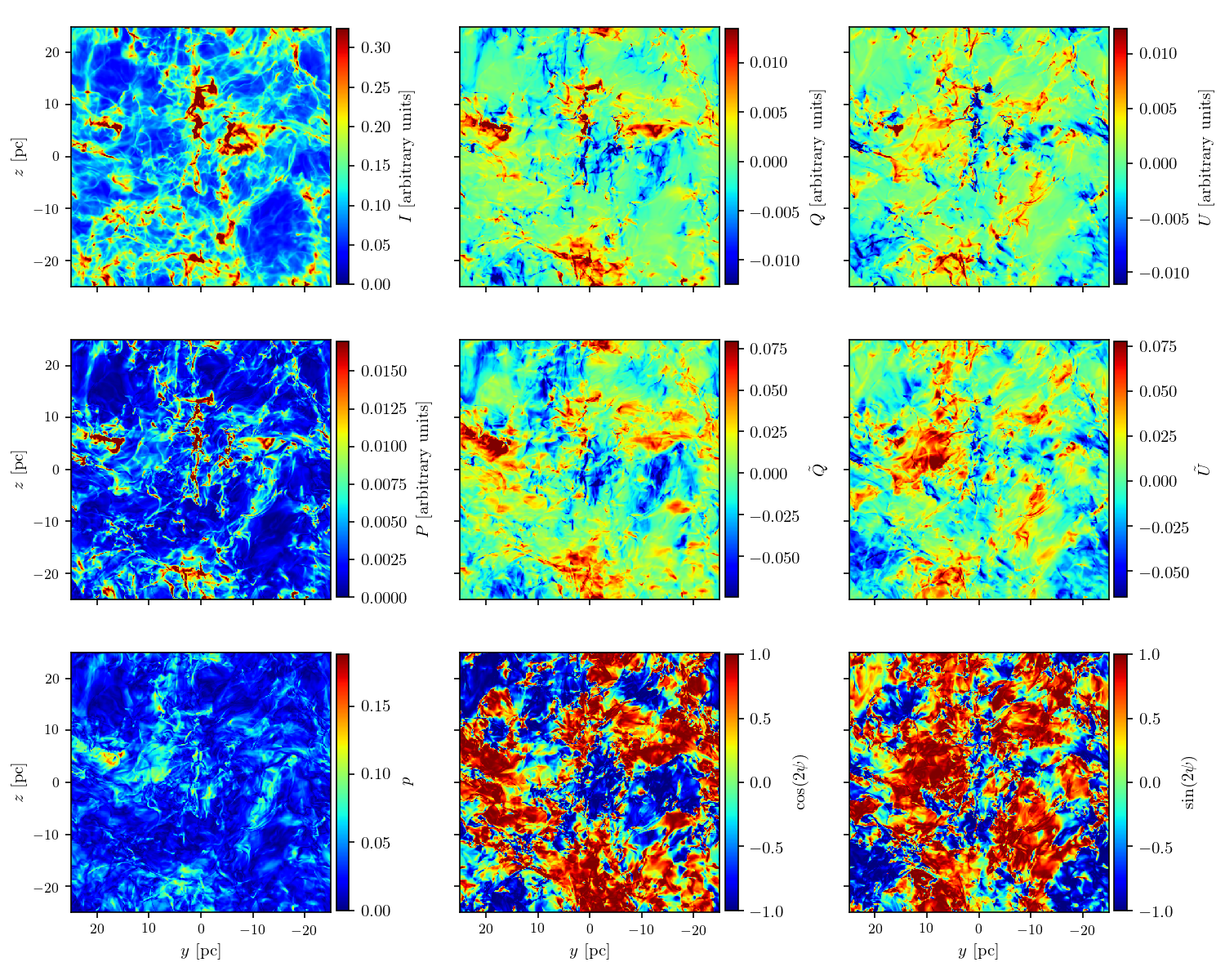}
	\caption{Same as Fig.~\ref{datasetOrtho} but for the $I_{\parallel}$, $Q_{\parallel}$, $U_{\parallel}$, $P_{\parallel}$, $\tilde{Q}_{\parallel}$, $\tilde{U}_{\parallel}$, $p_{\parallel}$, $\cos(2\psi)_{\parallel}$, and $\sin(2\psi)_{\parallel}$ data sets.}
	\label{datasetParallel}
\end{figure*}

The WST coefficients of polarization variables $\tilde{Q}+i\tilde{U}$, $\exp(2i\psi)$, and $p$ define statistical descriptions of polarization maps.
We expect some form of regularity in these coefficients that should lead to possible simplifications. This was the case for total intensity maps, which led to defining a reduced set of coefficients~\citep{Allys2019}. In this section, we similarly define the reduced wavelet scattering transform (hereafter RWST) to characterize the statistics of polarization data.

We first present the simulated sets of polarization maps we built from a numerical simulation of the diffuse interstellar medium. Then, we show what kind of statistical regularity arises from the WST descriptions of these data sets. We finally define RWST descriptions of these maps which are based on the modeling of the angular dependencies of the WST coefficients.

\subsection{Building simulated Stokes maps}
\label{SimulationDescription}

Throughout this paper we work with Stokes maps which are directly computed as integrations along a given direction of 3D simulated cubes of data extracted from a numerical simulation of magnetized interstellar turbulence (this given direction of integration corresponds to the line of sight for observations). As a first step, working with MHD simulations instead of observational data is a stepping-stone to relate our statistical descriptions to physics where we avoid the difficulty of dealing with data noise.

We use a MHD simulation designed to study the biphasic nature of the diffuse interstellar medium (Bellomi et al., in prep.). The simulation uses the adaptive mesh refinement code {\tt RAMSES}~\citep{Teyssier2002,Fromang2006} to solve the equations of ideal MHD as described in \citet{Iffrig2017}, neglecting self-gravity and taking into account heating and cooling processes of the gas. Turbulent forcing is applied to inject kinetic energy and balance numerical dissipation so that the simulation reaches a statistical steady state. In practice this turbulent forcing consists in a large-scale stochastic force field (a three component Ornstein-Uhlenbeck process defined in spectral space). Details on the turbulent forcing model can be found in \citet{Schmidt2006}.

The simulation volume is a $(50~\mathrm{pc})^3$ box divided into $512^3$ cells with periodic boundary conditions. At $t=0$ the gas has uniform properties with a density $n_\mathrm{H}=1.5~\mathrm{cm}^{-3}$ and a temperature $T = 8000~\mathrm{K}$, and the magnetic field is also uniform, with $\ve{B}_0 = B_0\ve{e}_x$ and $B_0 \sim 3.8~\mathrm{\mu G}$. In steady state, the turbulent forcing leads to an approximate velocity dispersion $\sigma_v \sim 2.6~\mathrm{km/s}$. This gives a turnover time at large scale $\tau_L \approx 18.8~\mathrm{Myr}$. Finally, an isotropic Habing radiation field is applied at the boundaries of the box. Its intensity is scaled by a factor $G_0 = 1$ \citep[for a definition, see][]{Draine2011}.

Once the simulation has reached a statistical steady state, we extract 14 snapshots that are approximately statistically independent using an approach similar to~\citet{Federrath2009}. In practice we make sure that two consecutive snapshots correspond to a minimal time evolution of $\delta\tau = 1.25~\mathrm{Myr}$ which is roughly ten percent of $\tau_L$. The phenomenology of turbulence in the sense of Kolmogorov shows that the turnover time $\tau_l$ at a given scale $l$ scales as $l^{2/3}$~\citep{Frisch1995}. This scaling holds for incompressible hydrodynamical turbulence only but we assume for the sake of simplicity that it extends reasonably well to ideal compressible MHD. At the range of scales considered in the following, we find that $\delta\tau$ is about fives times smaller than the corresponding largest value of $\tau_l$. Although this value is not completely satisfactory, we assume that the snapshots are statistically independent.

We can compute a set of Stokes maps $I$, $Q$, and $U$ for each of these snapshots using Eqs.~(\ref{eqI}), (\ref{eqQ}), and (\ref{eqU}) and choosing the $z$-axis of the cubes as the direction of integration. We note that this integration procedure is identical to the one used in \citet{planck2014-XX}. We choose a typical value for the polarization fraction parameter $p_0 = 0.2$ and arbitrary values for $\sigma_\nu$ and $T_d$ as these only determine the global amplitude of $I$, $Q$ and $U$ maps but do not impact the analysis. These maps are relevant for any frequency $\nu$ provided that the dust emission remains optically thin. Then from these Stokes maps we compute the associated maps $P$, $\tilde{Q}$, $\tilde{U}$, $p$, $\cos(2\psi)$, and $\sin(2\psi)$ which are the transformations of the Stokes observables defined in Sect.~\ref{transfoStokes}. Finally, we end up with a set of 14 maps for each of these variables.

We point out that the initial conditions of the MHD simulations are anisotropic because of the initial direction of the uniform magnetic field. This anisotropy remains once the simulation has reached steady state, due to magnetic flux conservation, and the value of the mean magnetic field is $\bar{\ve{B}}\approx \ve{B}_0$. Because the direction of integration chosen to compute the previous Stokes maps is orthogonal to the direction of the mean magnetic field $\bar{\ve{B}}$ in the simulation, we refer to the previous sets of maps using the $\bot$ symbol. For instance, the \textit{$p_{\bot}$ data set} refers to the set of 14 maps of polarization fraction $p$ computed from the 14 maps $I$, $Q$, and $U$ for which the $z$-axis was the axis of integration. We similarly compute Stokes maps integrating along the $x$-axis which is the direction of the mean magnetic field. This results in a set of maps that are statistically isotropic and from which we also compute the associated maps $P$, $\tilde{Q}$, $\tilde{U}$, $p$, $\cos(2\psi)$, and $\sin(2\psi)$. We use the $\parallel$ symbol to refer to these sets of maps.

\subsection{Presentation of the data sets}

We use for this work eight data sets, each one comprising 14 statistically independent maps. The various maps presented in the last subsection define 6 data sets: $\exp(2i\psi)_\bot, p_\bot, \tilde{Q}_\bot + i\tilde{U}_\bot$, $\exp{(2i\psi)}_\parallel, p_\parallel$, and $\tilde{Q}_\parallel + i\tilde{U}_\parallel$. In addition, we also analyze phase randomized data sets $R[\tilde{Q}_\bot+i\tilde{U}_\bot]$ and $R[\tilde{Q}_\parallel+i\tilde{U}_\parallel]$ built respectively from the $\tilde{Q}_\bot+i\tilde{U}_\bot$ and $\tilde{Q}_\parallel+i\tilde{U}_\parallel$ data sets.

We define $R[\cdot]$ to be the operator that acts on a map by randomizing the phases of the map in Fourier space, meaning that the new phases are drawn from a uniform distribution on $[0, 2\pi)$. We note that the modulus of the Fourier coefficients are retained so that the power spectrum is also unchanged. The phase information of an image is tightly bound to its structure~\citep{Oppenheim1981} and the main effect of the $R[\cdot]$ operator is to severely damage the structure of the image. We use phase randomization as an approximation to Gaussianization\footnote{Stationary Gaussian random fields (GRFs) do have uniformly distributed phases on $[0, 2\pi)$ but this property alone does not define them \citep[see][]{Wandelt2013}.}. We could have used a standard Gaussian random field generation \citep[see e.g.,][]{Kroese2015}, however we found out that the two approaches give similar results. In practice $R[\tilde{Q}_\bot+i\tilde{U}_\bot]$ (respectively $R[\tilde{Q}_\parallel+i\tilde{U}_\parallel]$) refers to the set of 14 maps produced by randomizing separately $\tilde{Q}$ and $\tilde{U}$ maps from the $\tilde{Q}_\bot+i\tilde{U}_\bot$ data set (respectively $\tilde{Q}_\parallel+i\tilde{U}_\parallel$). Technical details about phase randomization can be found in Appendix~\ref{PhaseRandomization}, as well as an example of a phase randomized map for $R[\tilde{Q}_\bot+i\tilde{U}_\bot]$ in Fig.~\ref{datasetRandomized}.

Figure~\ref{datasetOrtho} shows examples of maps for $I_{\bot}$, $Q_{\bot}$, $U_{\bot}$, $P_{\bot}$, $\tilde{Q}_{\bot}$, $\tilde{U}_{\bot}$, $p_{\bot}$, $\cos(2\psi)_{\bot}$, and $\sin(2\psi)_{\bot}$ data sets, while Fig.~\ref{datasetParallel} shows examples of maps for the corresponding $\parallel$ data sets. These figures show maps that all rely on the same snapshot of a MHD simulation. We draw the attention of the reader to a few points. First, we clearly see filamentary patterns on these maps that will demand a statistical description involving higher orders statistics compared to simple power spectra. We also note that $P$ is an order of magnitude lower than the intensity $I$ (this is due to the value of the polarization fraction parameter $p_0$), hence we have approximately $\tilde{Q}+i\tilde{U} \approx (Q+iU)/I$ so that $p$ roughly behaves as the modulus of $\tilde{Q}+i\tilde{U}$, and $2\psi$ roughly behaves as its complex argument. Next, we note that the magnitude of $p_\parallel$ is unsurprisingly much lower than that of $p_{\bot}$ as the direction of the projection of the local magnetic field on the plane of the sky is much less coherent along the line of sight when the line of sight corresponds to the direction of the mean magnetic field. Finally, we see on the $\cos(2\psi)_\bot$ map that the anisotropy of the magnetic field in the simulation leads to values that are concentrated close to 1.

\begin{figure}
	\centering
	\includegraphics[width=\hsize]{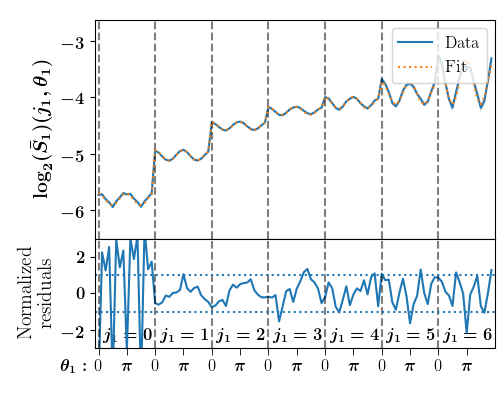}
	\caption{WST coefficients $\bar{S}_1(j_1, \theta_1)$ on a logarithmic scale for the $\tilde{Q}_{\bot}+i\tilde{U}_{\bot}$ data set, presented in a lexicographical order on $(j_1, \theta_1)$. Vertical dashed lines delimit distinct $j_1$ values. The top panel shows the original data (solid lines) and the RWST fit (dotted lines) corresponding to the model of Eq.~(\ref{eqRWSTS1}), while the bottom panel shows the normalized residuals of the fit.}
	\label{wstS1}
\end{figure}

\begin{figure*}
	\centering
	\includegraphics[width=\textwidth]{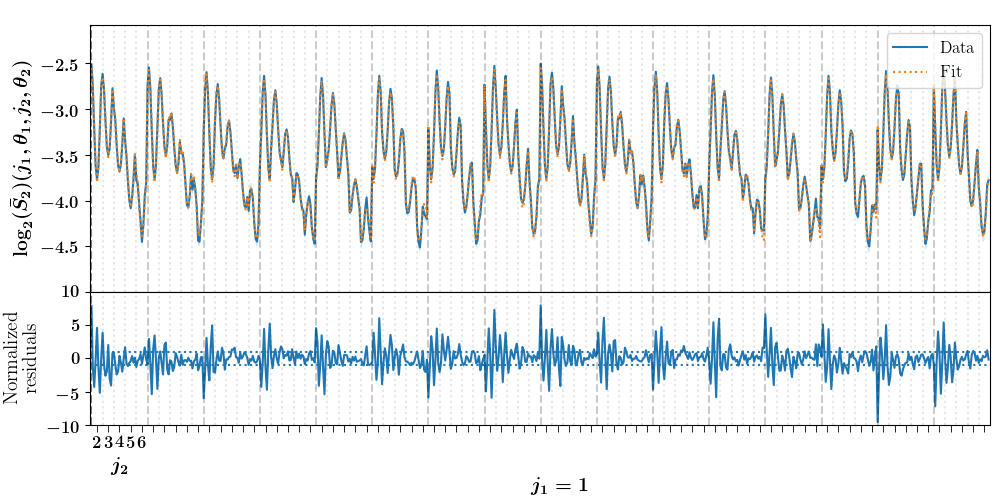}
	\caption{$j_1 = 1$ selection of WST coefficients $\bar{S}_2(j_1, \theta_1, j_2, \theta_2)$ on a logarithmic scale for the $\tilde{Q}_{\bot}+i\tilde{U}_{\bot}$ data set presented in a lexicographical order on $(j_1, \theta_1, j_2, \theta_2)$. This specific selection of coefficients is arbitrary, and we find similar results for other scales and the other data sets. Vertical dashed and dotted lines delimit distinct $\theta_1$ and $j_2$ values, respectively. The top panel shows the original data (solid lines) and the RWST fit (dotted lines) corresponding to the model of Eq.~(\ref{eqRWSTS2}), while the bottom panel shows the normalized residuals of the fit.}
	\label{wstS2}
\end{figure*}

\begin{figure}
	\centering
	\includegraphics[width=\hsize]{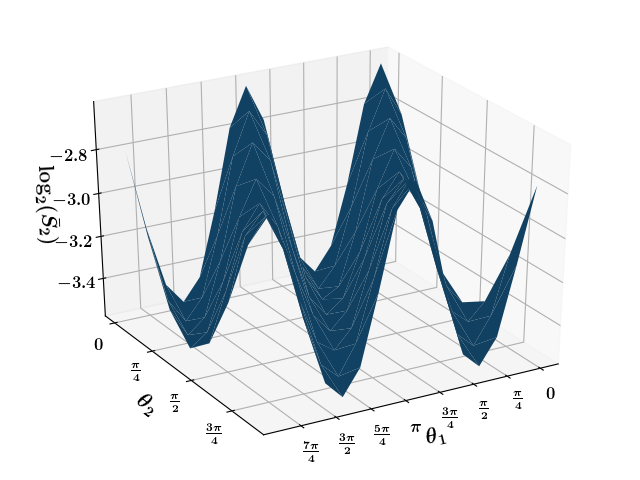}
	\caption{Surface representation of WST coefficients $\bar{S}_2(j_1=1, \theta_1,j_2=3,\theta_2)$ for the $\tilde{Q}_\bot+i\tilde{U}_\bot$ data set as a function of $\theta_1$ and $\theta_2$ variables only.}
	\label{wstS2AngularF}
\end{figure}

\subsection{Regularity in the WST coefficients}
\label{regularitySubsection}

We now focus on the WST coefficients associated with the ${\tilde{Q}_{\bot}+i\tilde{U}_{\bot}}$ data set, but the following reasoning remains valid for the other data sets, including those related to $\exp(2i\psi)$ and $p$.

\begin{table}
    \centering
    \begin{tabular}{|c|ccccccc|}
    \hline
    $j$ & 0 & 1 & 2 & 3 & 4 & 5 & 6 \\
    \hline
       $\tilde{\lambda}$ [pixels] & 2.67 & 5.33 & 10.7 & 21.3 & 42.7 & 85.3 & 171 \\ 
       $\lambda$ [pc] & 0.26 & 0.52 & 1.04 & 2.08 & 4.17 & 8.33 & 16.7 \\
    \hline
    \end{tabular}
    \caption{Correspondence between scale index $j$ and related wavelengths on simulated maps, both in pixel units and dimensional units. Those wavelengths $\tilde{\lambda}$ come from the definition of dilated Morlet wavelets $\psi_{j, \theta}$~\citep[for more details, see][]{Bruna2013}.}
    \label{tableScales}
\end{table}

For the 14 maps of the $\tilde{Q}_{\bot}+i\tilde{U}_{\bot}$ data set, we compute the WST coefficients for $J = 7$ and $\Theta = 8$ (see Eqs.~(\ref{S0eqCplx}), (\ref{S1eqCplx}), and (\ref{S2eqCplx})). $J$ could have been fixed to a higher value for $512\times512$ maps, but because of our limited number of maps we chose to restrict our statistical description to scales for which we have a sufficient number of modes for reliable estimations. We give in Table~\ref{tableScales} the correspondence between the scale index $j$ and the central wavelength of the related dilated Morlet wavelet both in pixel units and in dimensional units (related to the MHD simulation).

These $J$ and $\Theta$ values lead for each map to 112 $\bar{S}_1$ coefficients and 2688 $\bar{S}_2$ coefficients. Adding to this the $\bar{S}_0$ coefficient, we end up with 2801 coefficients per map. We define mean $\bar{S}_0$, $\bar{S}_1$ and $\bar{S}_2$ coefficients as means of the WST coefficients over the 14 maps and we also compute the standard deviation of the mean for each of these coefficients assuming that the maps are statistically independent.

Figures \ref{wstS1} and \ref{wstS2} represent (in blue) respectively $\bar{S}_1(j_1,\theta_1)$ coefficients and a representative subsample of $\bar{S}_2(j_1, \theta_1, j_2, \theta_2)$ coefficients (for $j_1 = 1$) on a logarithmic scale, for $\tilde{Q}_{\bot}+i\tilde{U}_{\bot}$. In both figures, we represent the multivariate functions $\bar{S}_1$ and $\bar{S}_2$ in a lexicographical order for the $(j_1, \theta_1)$ and $(j_1, \theta_1, j_2, \theta_2)$ variables, respectively. Vertical gray lines help to mark increments of these variables.

In Fig.~\ref{wstS1}, we see that for a fixed scale $j_1$ a smooth pattern emerges with respect to the angular variable $\theta_1$. Because of the definition of $\bar{S}_1$ coefficients, at fixed $j_1$, $\log_2(\bar{S}_1)(j_1,\cdot)$ functions must be $2\pi$-periodic, but here the smooth pattern appears to be $\pi$-periodic. While it is not surprising to get smooth angular patterns that reflect the regularity of physical processes, this $\pi$-periodicity with respect to the angular variable $\theta_1$ was unexpected.

We can nevertheless explain it by noticing that:
\begin{equation}
    |(\tilde{Q}+i\tilde{U})\star\psi_{j,\theta}| = |[(\tilde{Q}+i\tilde{U})\star\psi_{j,\theta}]^*| = |(\tilde{Q}-i\tilde{U})\star\psi_{j,\theta+\pi}|,
\end{equation}
because for Morlet wavelets we have $\psi_{j,\theta}^*=\psi_{j, \theta+\pi}$\footnote{The $^*$ symbol stands for the complex conjugate.}. Hence, saying that $\log_2(\bar{S}_1)(j_1,\cdot)$ functions are $\pi$-periodic amounts to saying that $\tilde{Q}+i\tilde{U}$ and $\tilde{Q}-i\tilde{U}$ have the same WST first order statistics.

We can identify the same kind of regularity properties for $\bar{S}_2$ coefficients in Fig.~\ref{wstS2}. Since $\bar{S}_2$ coefficients depend on four variables, two of which are angular variables $\theta_1$ and $\theta_2$, it is more complicated to see this angular regularity, but Fig.~\ref{wstS2AngularF} helps us to exhibit this regularity by plotting $\bar{S}_2$ coefficients as a function of angular variables $\theta_1$ and $\theta_2$ when $j_1 = 1$ and $j_2 = 3$. Smooth angular patterns appear on this surface which is parameterized by $\theta_1$ and $\theta_2$. In particular we see that $\theta_1 - \theta_2 = c$ cuts of this surface for arbitrary constants $c$ give roughly constant $\bar{S}_2$ coefficients. On this example we thus expect most of the angular modulation to depend on the $\theta_2 - \theta_1$ variable.

\subsection{Definition of a RWST statistical description}
\label{RWSTsubsection}

The smooth periodic patterns identified in the $\tilde{Q}_\bot + i\tilde{U}_\bot$ WST coefficients suggest that a simplification of the WST statistical description is possible, through an adequate modeling of its angular dependencies. The purpose of such a modeling of the WST coefficients is twofold: 1) lowering the dimensionality of the statistical description of our data, and 2) providing an interpretable representation of these angular dependencies in terms of considerations of isotropy and anisotropy of the data.

We model the regularity of WST coefficients with respect to angular variables $\theta_1$ and $\theta_2$ using the RWST model introduced in \citet{Allys2019}. It is remarkable that a model developed for total intensity maps may be applied to $\tilde{Q} + i\tilde{U}$ complex Stokes maps of polarized thermal emission of dust, without any modification. This highlights the generality of such an angular modeling of the WST coefficients for maps of dust emission, and this model may surely be extended to other kinds of complex-valued fields in physics. We present in Appendix~\ref{rwstGenerality} the RWST model in terms of Fourier series expansions and rephrase the characteristics of this generality.

We now briefly recall the RWST model, and refer the reader to \cite{Allys2019} for more details.

In the RWST model, the $\bar{S}_1$ coefficients are written as:
\begin{align}
    \label{eqRWSTS1}
    \log_2\left[\bar{S}_1\left(j_1, \theta_1\right)\right] = &\hat{S}_1^{\mathrm{iso}}(j_1) \nonumber \\ + &\hat{S}_1^{\mathrm{aniso}}(j_1)\cos\left(2\left[\theta_1 - \theta^{\mathrm{ref}, 1}\left(j_1\right)\right]\right),
\end{align}
where $\hat{S}_1^{\mathrm{iso}}(j_1)$, $\hat{S}_1^{\mathrm{aniso}}(j_1)$, and $\theta^{\mathrm{ref}, 1}(j_1)$ are the parameters of this angular model for scale $j_1$. This model thus depends on $3\times J$ parameters. We also enforce $\hat{S}_1^{\mathrm{aniso}}(j_1) \geq 0$ in order to lift a degeneracy between the $\hat{S}_1^{\mathrm{aniso}}(j_1)$ and $\theta^{\mathrm{ref}, 1}$ parameters. $\hat{S}_1^{\mathrm{iso}}(j_1)$ quantifies the isotropic component of the data fluctuations at a scale $j_1$, while $\hat{S}_1^{\mathrm{aniso}}(j_1)$ defines a measure of the angular modulation of the coefficients at scale $j_1$, introducing a reference angle $\theta^{\mathrm{ref}, 1}(j_1)$ that defines a privileged direction in the maps \footnote{We note that Eq.~(\ref{eqRWSTS1}) defines an angular model of the logarithm of the WST coefficients that turns angular modulations of the WST coefficients into additive corrections $\hat{S}_1^{\mathrm{aniso}}(j_1)\cos\left(2\left[\theta_1 - \theta^{\mathrm{ref}, 1}\left(j_1\right)\right]\right)$ to the isotropic amplitude of fluctuations $\hat{S}_1^{\mathrm{iso}}(j_1)$. We use a base 2 logarithm to be consistent with the base 2 definition of scales $j_1$ and $j_2$.}.

Similarly, the RWST model for $\bar{S}_2$ coefficients reads:
\begin{align}
     \label{eqRWSTS2}
       \log_2  & \left[ \bar{S}_2 \left(j_1, \theta_1, j_2, \theta_2\right)\right] =\hat{S}_2^{\mathrm{iso}, 1}(j_1, j_2) \nonumber \\ + & \hat{S}_2^{\mathrm{iso}, 2}(j_1, j_2)\cos\left(2\left[\theta_1 - \theta_2\right]\right)\nonumber \\+ & \hat{S}_2^{\mathrm{aniso},1}(j_1, j_2)\cos\left(2\left[\theta_1 - \theta^{\mathrm{ref}, 2}\left(j_1, j_2\right)\right]\right)\nonumber \\ + & \hat{S}_2^{\mathrm{aniso},2}(j_1, j_2)\cos\left(2\left[\theta_2- \theta^{\mathrm{ref}, 2}\left(j_1, j_2\right)\right]\right),
\end{align}
where $\hat{S}_2^{\mathrm{iso}, 1}(j_1,j_2)$, $\hat{S}_2^{\mathrm{iso}, 2}(j_1, j_2)$, $\hat{S}_2^{\mathrm{aniso},1}(j_1,j_2)$, $\hat{S}_2^{\mathrm{aniso},2}(j_1,j_2)$, and $\theta^{\mathrm{ref}, 2}(j_1,j_2)$ are the parameters of this angular model for each pair of scales $(j_1,j_2)$. As we have $j_2 > j_1$, we end up with $5\times J\times(J-1)/2$ parameters for this model. Here again we make sure that $\hat{S}_2^{\mathrm{aniso}, 1}(j_1, j_2) \geq 0$ to avoid any parameter degeneracy. $\hat{S}_2^{\mathrm{iso}, 1}$ measures the overall amplitude of coupling between the scales $j_1$ and $j_2$. $\hat{S}_2^{\mathrm{iso}, 2}$ represents the amplitude of the modulation due to the relative orientation of the wavelets $\psi_{j_1, \theta_1}$ and $\psi_{j_2, \theta_2}$, and we interpret it as a signature of filamentary structures at a given scale. Indeed for an oriented filamentary structure, we expect the $\bar{S}_2$ coefficients to peak when $\theta_2 = \theta_1$ and to reach a minimum when $\theta_2 = \theta_1 + \pi/2$. Finally, $\hat{S}_2^{\mathrm{aniso},1}$ and $\hat{S}_2^{\mathrm{aniso},2}$ are measures of the anisotropic properties of the data in second order WST coefficients, here decoupling $\theta_1$ and $\theta_2$ contributions.

In practice, for a given data set, this RWST model of the angular dependencies is fitted to the first order WST coefficients for every scale $j_1$, and to the second order WST coefficients for every pair of scales $(j_1, j_2)$ (with $j_2 > j_1$). The accuracy of these multiple fits is quantified with $\chi_\mathrm{r}^2$ statistics as described in \citet{Allys2019}. Since it is not possible to properly estimate the full covariance matrix with only 14 samples per coefficient, the uncertainties affecting the WST coefficients for a given data set are simply estimated from the sample variance across the various simulation snapshots.

However an important correlation of the first order WST coefficients across angles for each scale $j_1$ needs to be taken into account to properly estimate statistical uncertainties. For each sample we compute a mean coefficient across angles for a given scale and subtract this mean before computing the statistical uncertainties\footnote{This decreases the effective number of degrees of freedom by one when computing $\chi_\mathrm{r}^2$ values.}. This mitigates most of the correlation between WST coefficients at the same scale $j_1$.

\begin{figure}
	\centering
	\includegraphics[width=0.9\hsize]{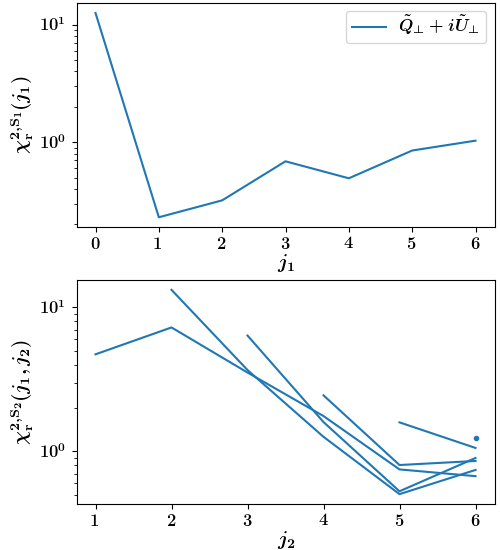}
	\caption{Reduced chi square $\chi_\mathrm{r}^{2, \mathrm{S_1}}(j_1)$ (top) and $\chi_\mathrm{r}^{2, \mathrm{S_2}}(j_1, j_2)$ (bottom) associated with the RWST fit of the WST coefficients (see Eqs.~(\ref{eqRWSTS1}) and (\ref{eqRWSTS2})) for the $\tilde{Q}_\bot+i\tilde{U}_\bot$ data set. In the bottom panel, each curve corresponds to a fixed $j_1$ value with $j_2$ varying from $j_1 + 1$ to $J-1=6$. For $j_1 = J - 2$, the curve is reduced to a single dot on the figure. We use logarithmic scales for better visibility.}
	\label{rwstChi2r}
\end{figure}

Figure ~\ref{rwstChi2r} shows the $\chi_\mathrm{r}^2$ values for both $\log_2(\bar{S}_1)$ and $\log_2(\bar{S}_2)$ fits for the $\tilde{Q}_\bot+i\tilde{U}_\bot$ data set. The $\chi^2_\mathrm{r}$ values are globally close to 1 except at low $j_1$ for $\chi_\mathrm{r}^{2, \mathrm{S_1}}$ and at low $j_2 - j_1$ for $\chi_\mathrm{r}^{2, \mathrm{S_2}}$. This deterioration of the goodnesses of the fits is due to a pixellization effect at small scales, and may be fixed by adding adequate lattice terms in the RWST model, as described in the appendix C of \citet{Allys2019}. This is shown for our data in Appendix~\ref{LatticeCorrections}. We note that these additional terms do not affect the values of the coefficients corresponding to the original RWST model. Finally, the same RWST model applies to the other data sets used in this work, including $\exp(2i\psi)$ and $p$ data sets, and we get for all of them similar $\chi_\mathrm{r}^2$ values.

In Figs.~ \ref{wstS1} and \ref{wstS2} we show the RWST fit overplotted on a selection of first and second order WST coefficients, and also show the corresponding normalized residuals. The curves are globally in good agreement and the flaws of the RWST model due to numerical effects at low $j_1$ and $j_2 - j_1$ appear as clear patterns in the residuals. Additional figures in appendix \ref{LatticeCorrections} show how these artifacts may be taken into account.

In the end, the RWST coefficients define statistical descriptions of the data sets in terms of simple considerations of isotropy and anisotropy with respect to a reference direction. Furthermore, these statistical descriptions exhibit much lower dimensionality, with a total of 127 coefficients (including $\bar{S}_0$ coefficient in the description) compared to the original 2801 WST coefficients in the case of complex variables $\tilde{Q}+i\tilde{U}$ and $\exp(2i\psi)$\footnote{For $p$ maps, the WST descriptions consist of 1401 coefficients.}, thus providing a very large compression of the statistical information contained in the WST coefficients.

%%%%%%%%%%%%%%%%%%%%%%%%%%%%%%%%%%%%%%%%%%%%%%%%%%%%
%%%%%%%%%%%%%%%%%%%%%%%%%%%%%%%%%%%%%%%%%%%%%%%%%%%%
%% SECTION 4: 
%%%%%%%%%%%%%%%%%%%%%%%%%%%%%%%%%%%%%%%%%%%%%%%%%%%%
%%%%%%%%%%%%%%%%%%%%%%%%%%%%%%%%%%%%%%%%%%%%%%%%%%%%

\section{RWST data analysis and interpretation}
\label{section4}

In this section we analyze and interpret the RWST statistical descriptions for the 8 data sets built from the same MHD simulation. We relate the coefficients of these descriptions to the physical properties of the simulation.

\subsection{Isotropic fluctuations in first order coefficients $\hat{S}_1^{\mathrm{iso}}$}
\label{S1IsoSubsection}

\begin{figure}
	\centering
	\includegraphics[width=0.9\hsize]{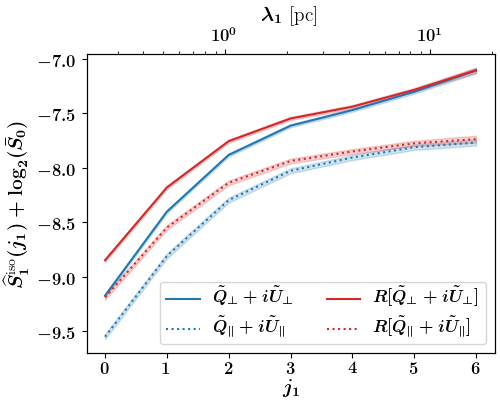}
	\caption{$\hat{S}_1^{\mathrm{iso}}(j_1)+\log_2(\bar{S}_0)$ RWST coefficients for $\tilde{Q}_\bot+i\tilde{U}_\bot, \tilde{Q}_\parallel+i\tilde{U}_\parallel$, $R[\tilde{Q}_\bot+i\tilde{U}_\bot]$ and $R[\tilde{Q}_\parallel+i\tilde{U}_\parallel]$ data sets. We use solid lines for $\bot$ data sets and dotted lines for $\parallel$ data sets. For reference, we show at the top of this figure the correspondence between $j$ scale indices and the related wavelengths on the maps (in practice, we have $\lambda_1 = 2^{j_1+1}\pi/k_0$).}
	\label{S1IsoQiU}
\end{figure}

Figure~\ref{S1IsoQiU} presents $\hat{S}_1^{\mathrm{iso}} + \log_2(\bar{S}_0)$ coefficients as a function of scale for $\tilde{Q}_\bot+i\tilde{U}_\bot, \tilde{Q}_\parallel+i\tilde{U}_\parallel$, $R[\tilde{Q}_\bot+i\tilde{U}_\bot]$ and $R[\tilde{Q}_\parallel+i\tilde{U}_\parallel]$ data sets. We note that the $\log_2(\bar{S}_0)$ term cancels the normalization of the $\bar{S}_1$ WST coefficients defined in Eq.~(\ref{S1eqCplx}) in order to analyze the statistics of $\tilde{Q}+i\tilde{U}$ (and not the statistics of $(\tilde{Q}+i\tilde{U})/\langle|\tilde{Q}+i\tilde{U}|\rangle$). We see that $\hat{S}_1^{\mathrm{iso}} + \log_2(\bar{S}_0)$ levels which measure an amplitude of fluctuations of the signal, per scale, are higher for $\tilde{Q}_\bot+i\tilde{U}_\bot$ and $R[\tilde{Q}_\bot+i\tilde{U}_\bot]$ than for the corresponding $\parallel$ data sets. This shows that the amplitudes differ depending on the orientation of the mean magnetic field with respect to the line of sight. We have more fluctuations within $\tilde{Q}+i\tilde{U}$ maps when the mean magnetic field is in the plane of the sky.

The differences between the $\tilde{Q}+i\tilde{U}$ data set and its corresponding randomized counterpart $R[\tilde{Q}+i\tilde{U}]$ (in both $\bot$ and $\parallel$ cases) illustrate the difference between $\bar{S}_1$ coefficients and the power spectrum. Indeed, $\tilde{Q}+i\tilde{U}$ and $R[\tilde{Q}+i\tilde{U}]$ maps share the same power spectrum but have different $\bar{S}_1$ coefficients. The fact that the $R[\cdot]$ operator increases the $\bar{S}_1$ values leaving the power spectrum unchanged shows that it reduces the sparsity of these maps (see discussion in Sect.~\ref{wstSubsection}). This feature underlines the non-Gaussianity of the $\tilde{Q}+i\tilde{U}$ data set and we therefore expect its second order coefficients to be higher compared to those of the corresponding randomized data set. We note that these differences wear off at large scales. We interpret this as statistical evidence that the non-Gaussianity of $\tilde{Q}+i\tilde{U}$ maps decreases at large scales. This result may reflect a characteristic of interstellar turbulence but could also follow from the fact that we start to probe the Gaussian distribution of the turbulent forcing of the simulation.

\begin{figure}
	\centering
	\includegraphics[width=0.9\hsize]{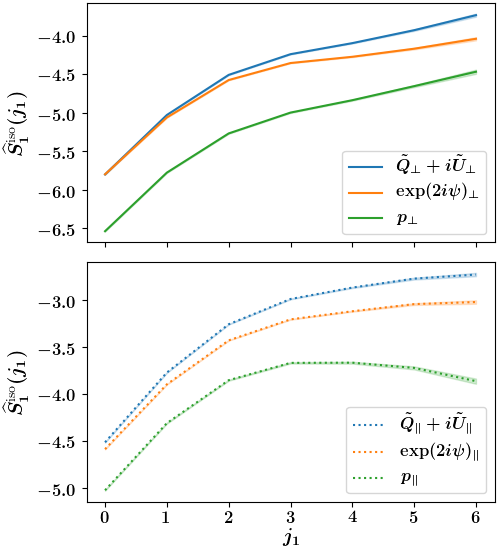}
	\caption{$\hat{S}_1^{\mathrm{iso}}(j_1)$ RWST coefficients for $\tilde{Q}+i\tilde{U}, \exp(2i\psi)$ and $p$ data sets in the $\bot$ (top, solid lines) and $\parallel$ (bottom, dotted lines) cases.}
	\label{S1Iso}
\end{figure}

In Fig.~\ref{S1Iso}, we compare $\tilde{Q}+i\tilde{U}$ $\hat{S}_1^{\mathrm{iso}}$ coefficients to those of $p$ and $\exp(2i\psi)$ for respectively $\parallel$ and $\bot$ data sets. Since we have $\tilde{Q}+i\tilde{U} \approx p\exp(2i\psi)$ we would like to compare the relative contributions of $\hat{S}_1^{\mathrm{iso}}$ coefficients of $\tilde{Q}+i\tilde{U}$ maps, but this seems more complicated than expected as we found out that a proper comparison through $\hat{S}_1^{\mathrm{iso}}$ coefficients highly depends on the choice of normalization of the WST coefficients for $p$.

$\hat{S}_1^{\mathrm{iso}}$ coefficients roughly range from $-4.5$ to $-3.0$ for $\exp({2i\psi})_\parallel$ while they range from $-5.8$ to $-4.2$ for $\exp({2i\psi})_\bot$. These differences indicate larger fluctuations of the polarization angle when the mean magnetic field is along the line of sight compared to when the mean magnetic field is in the plane of the sky. Since the average polarization fraction $p$ is lower when the mean magnetic field is along the line of sight ($\bar{S}_0 \approx 0.03 $ for $p_\parallel$ compared to $\bar{S}_0 \approx 0.1$ for $p_\bot$) this feature is consistent with the anti-correlation observed with polarization data between the angle dispersion function $\mathcal{S}$ and the polarization fraction $p$~\citep{planck2014-XX,planck2016-l11B,Fissel2016}.

\subsection{Anisotropic fluctuations in first order coefficients $\hat{S}_1^{\mathrm{aniso}}$}

\begin{figure}
	\centering
	\includegraphics[width=0.9\hsize]{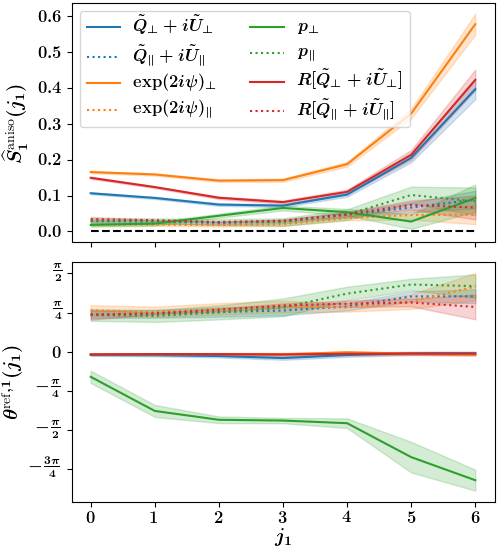}
	\caption{$\hat{S}_1^{\mathrm{aniso}}(j_1)$ (top) and $\theta^{\mathrm{ref},1}(j_1)$ (bottom) RWST coefficients for $\tilde{Q}+i\tilde{U}, \exp(2i\psi)$, $p$, and $R[\tilde{Q}+i\tilde{U}]$ data sets in the $\bot$ (solid lines) and $\parallel$ (dotted lines) cases.}
	\label{S1Aniso}
\end{figure}

$\hat{S}_1^{\mathrm{aniso}}$ coefficients measure the angular modulation of the first order WST coefficients. They are presented in Fig.~\ref{S1Aniso} for all data sets. We see that this anisotropy is much larger for $\bot$ data sets than for $\parallel$ ones. The $p_\bot$ data set constitutes an exception among the $\bot$ data sets as it has a rather low anisotropy level. These differences are not surprising as we expect a stronger anisotropy when the mean magnetic field is in the plane of the sky, while statistical isotropy is expected when integrating along the mean magnetic field. For larger scales, we see an increase of these coefficients for $\bot$ data sets. This trend has already been observed on observational data of the Polaris flare in total intensity~\citep{Allys2019} and deserves a closer examination.

As explained in Sect.~\ref{SimulationDescription} the large scales of consecutive snapshots could be correlated to some extent. To assess the potential impact of these correlations on our analysis, we have computed separate RWST statistics for three maps of the ${\tilde{Q}_\bot+i\tilde{U}_\bot}$ data set corresponding to snapshots that are sufficiently distant in time to rightfully assume the independence (we choose $6\times\delta\tau$ instead of $\delta\tau$). The increase of $\hat{S}_1^{\mathrm{aniso}}$ coefficients remains significant for each map, which demonstrates that this trend is not a consequence of correlations among snapshots.

Reference angles $\theta^{\mathrm{ref}, 1}$ also presented in Fig.~\ref{S1Aniso} show that the preferential direction identified for anisotropic $\bot$ data sets is the direction corresponding to $\theta^{\mathrm{ref}, 1} = 0$ for all scales $j_1$. Such a value of the reference angle indicates a statistical tendency for structures, including filaments, to be elongated vertically rather than horizontally, that is, along the $y$ axis in Fig.~\ref{datasetOrtho}. This corresponds to an elongation which is orthogonal to the mean magnetic field. This result is to be compared in further works to the abundant literature on the relative orientation between magnetic fields and structures traced by interstellar dust~\citep[for a review, see][]{Hennebelle2019}.

The reference angle found for $\parallel$ data set is well defined and approximately equal to $\pi/4$ for all scales while the anisotropy levels are close to zero. These values of $\theta^{\mathrm{ref}, 1}$ are surprising because we were not expecting any anisotropy for these data sets. By examining the RWST statistics separately for each map, for each of the corresponding data sets, we found out that these surprising values correspond to an intermittent rise of the anisotropy level that appears in a few consecutive snapshots of the simulation. In this case where the level and direction of anisotropy are not coherent over snapshots, the mean coefficient gives an incomplete view of the anisotropic properties of the simulation. Notably, even with significant levels of anisotropy per map, if the reference angles are incoherent, we expect the mean level of anisotropy to be small. We found out that this is what happens for the $p_\bot$ data set, where $\hat{S}_1^{\mathrm{aniso}}$ coefficients are relatively small while $\theta^{\mathrm{ref}, 1}$ coefficients clearly deviate from zero.

\subsection{Second order coefficients and non-Gaussianity of the data}

\begin{figure}
	\centering
	\includegraphics[width=0.9\hsize]{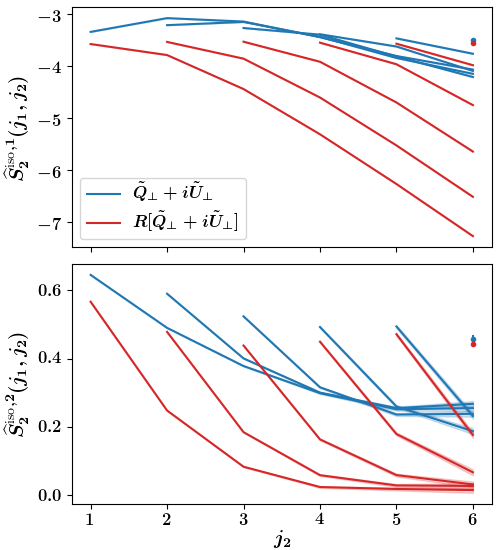}
	\caption{$\hat{S}_2^{\mathrm{iso}, 1}(j_1,j_2)$ (top) and $\hat{S}_2^{\mathrm{iso}, 2}(j_1,j_2)$ (bottom) RWST coefficients for $\tilde{Q}_\bot+i\tilde{U}_\bot$ and $R[\tilde{Q}_\bot+i\tilde{U}_\bot]$ data sets. Each curve corresponds to a fixed $j_1$ value with $j_2$ varying from $j_1 + 1$ to $J-1=6$. For $j_1 = J - 2$, the curve is reduced to a single dot on the figure.}
	\label{S2IsoQiU}
\end{figure}

\begin{figure*}
	\centering
	\includegraphics[width=\textwidth]{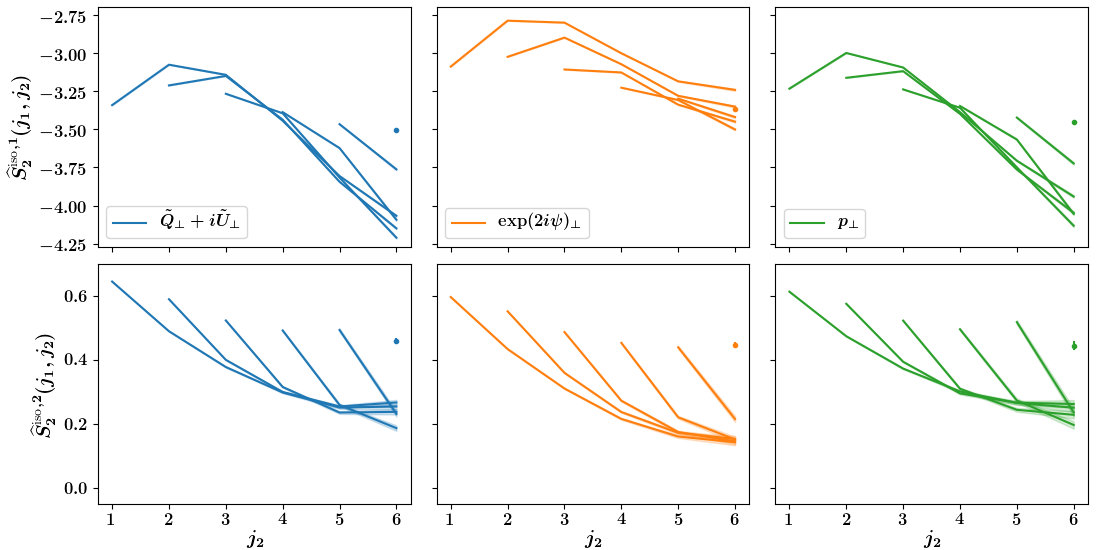}
	\caption{$\hat{S}_2^{\mathrm{iso}, 1}(j_1,j_2)$ (left column) and $\hat{S}_2^{\mathrm{iso}, 2}(j_1,j_2)$ (right column) RWST coefficients for $\tilde{Q}+i\tilde{U}$, $\exp(2i\psi)$ and $p$ data sets in the $\bot$ case. Each curve corresponds to a fixed $j_1$ value with $j_2$ varying from $j_1 + 1$ to $J-1=6$. For $j_1 = J - 2$, the curve is reduced to a single dot on the figure.}
	\label{S2Iso}
\end{figure*}

$R[\tilde{Q}+i\tilde{U}]$ maps are Gaussian approximations of $\tilde{Q}+i\tilde{U}$ maps, and we have already exhibited differences between these data sets in their first order RWST coefficients in Sect.~\ref{S1IsoSubsection}. Similarly, second order RWST coefficients show remarkable differences that highlight the non-Gaussianity of the $\tilde{Q}_\bot+i\tilde{U}_\bot$ and $\tilde{Q}_\parallel+i\tilde{U}_\parallel$ data sets. The two dominant second order RWST coefficients $\hat{S}_2^{\mathrm{iso}, 1}$ and $\hat{S}_2^{\mathrm{iso}, 2}$ presented in Fig.~\ref{S2IsoQiU} display clearly distinct patterns between the original data sets and the randomized ones on the example of the $\bot$ data sets. First, $\hat{S}_2^{\mathrm{iso}, 1}$ and $\hat{S}_2^{\mathrm{iso}, 2}$ coefficients are globally smaller for the $R[\tilde{Q}+i\tilde{U}]$ data sets, which is in line with what we had foreseen in Sect.~\ref{S1IsoSubsection}. In addition, $\hat{S}_2^{\mathrm{iso}, 1}$ coefficients for $R[\tilde{Q}+i\tilde{U}]$ show a scale invariance property: $\hat{S}_2^{\mathrm{iso}, 1}$ coefficients only depend on the difference $j_2 - j_1$. We point out that these scale invariant patterns are signatures of scale invariant Gaussian processes~\citep{Bruna2015} and have already been observed for fractional Brownian motions processes in~\citet{Allys2019}.

$\hat{S}_2^{\mathrm{iso},2}$ coefficients also show two distinct trends: the coefficients quickly tend to zero when $j_2 - j_1$ increases for $R[\tilde{Q}+i\tilde{U}]$ while coefficients tend to strictly positive values for the largest $j_2 - j_1$ values for non randomized data sets $\tilde{Q}+i\tilde{U}$. This result is related to the filamentary structure of the non randomized maps, because a filamentary structure involves a modulation of the WST coefficients as a function of the angle difference $\theta_2 - \theta_1$. We see the same trend in Fig.~\ref{S2Iso} for the $\hat{S}_2^{\mathrm{iso},2}$ coefficients of the other non randomized data sets which all present filamentary structures. For all data sets, and fixed $j_1$ values, $\hat{S}_2^{\mathrm{iso},2}$ coefficients decrease as a function of $j_2$. This property seems to be general. We also expect a signal due to the imprint of the wavelets in the $\hat{S}_2^{\mathrm{iso},2}$ coefficients that would increase their values when $j_2$ is close to $j_1 + 1$.

We see in Fig.~\ref{S2IsoQiU} that the differences between the $\tilde{Q}_\bot+i\tilde{U}_\bot$ data set and its randomized counterpart decrease for the highest $j_1$ values. Here again, this shows that the non-Gaussianity of ${\tilde{Q}+i\tilde{U}}$ maps decreases at large scales. This is consistent with what we have already pointed out in Sect.~\ref{S1IsoSubsection} for $\hat{S}_1^{\mathrm{iso}}$ coefficients and we interpret this trend similarly.

Second order anisotropic coefficients $\hat{S}_2^{\mathrm{aniso},1}, \hat{S}_2^{\mathrm{aniso},2}$, and $\theta^{\mathrm{ref},2}$ essentially show consistent results with first order anisotropic coefficients both in terms of amplitude and direction of anisotropy. They have generally smaller values compared to those of isotropic coefficients $\hat{S}_2^{\mathrm{iso},1}$ and $\hat{S}_2^{\mathrm{iso},2}$. We do not show these coefficients here as we do not discuss them any further in this work.

%%%%%%%%%%%%%%%%%%%%%%%%%%%%%%%%%%%%%%%%%%%%%%%%%%%%
%%%%%%%%%%%%%%%%%%%%%%%%%%%%%%%%%%%%%%%%%%%%%%%%%%%%
%% SECTION 5: 
%%%%%%%%%%%%%%%%%%%%%%%%%%%%%%%%%%%%%%%%%%%%%%%%%%%%
%%%%%%%%%%%%%%%%%%%%%%%%%%%%%%%%%%%%%%%%%%%%%%%%%%%%

\section{Random syntheses of $\tilde{Q}+i\tilde{U}$ polarization maps}
\label{SectionSyntheses}

The RWST coefficients are not only statistical descriptors of the data but can also be used as a set of constraints for generating multiple random realizations of polarization maps. By comparing the original and synthetic maps one can assess the exhaustiveness of the statistical description. A visual agreement would give more confidence in the relevance of the statistical information captured by a RWST description. However, this qualitative assessment has its limits as it is unclear what kind of statistics human eyes are sensitive to~\citep{Julesz1981}. A quantitative comparison using summary statistics other than the RWST is thus also needed.

The synthesis is also a means of simulating noise-free maps artificially and realistically augmenting demanding MHD simulations, or producing realistic foregrounds for component separation methods for CMB data analysis. This approach differs from previous work where dust polarization maps are computed from a phenomenological model and ancillary observations \citep{planck2016-XLII,Ghosh2017,Vansyngel2017,Levrier2018,Adak19,Clark19} or directly from numerical simulations \citep{Kim19}. It follows what has been done for dust total intensity by \citet{Allys2019} using the RWST, and \citet{Aylor19} using a Generative Adversarial Network.

In this section, we present random synthetic $\tilde{Q}+i\tilde{U}$ maps generated from RWST coefficients derived from the analysis of our MHD simulation and complemented by additional constraints on the large-scale components of the maps and on the one-point probability distribution function. We then compare synthetic and original maps using one-point and two-point statistics.

\subsection{Generation of synthetic polarization maps from a RWST description}
\label{synthesesProcedure}

\begin{figure*}
	\centering
	\includegraphics[width=0.7\textwidth]{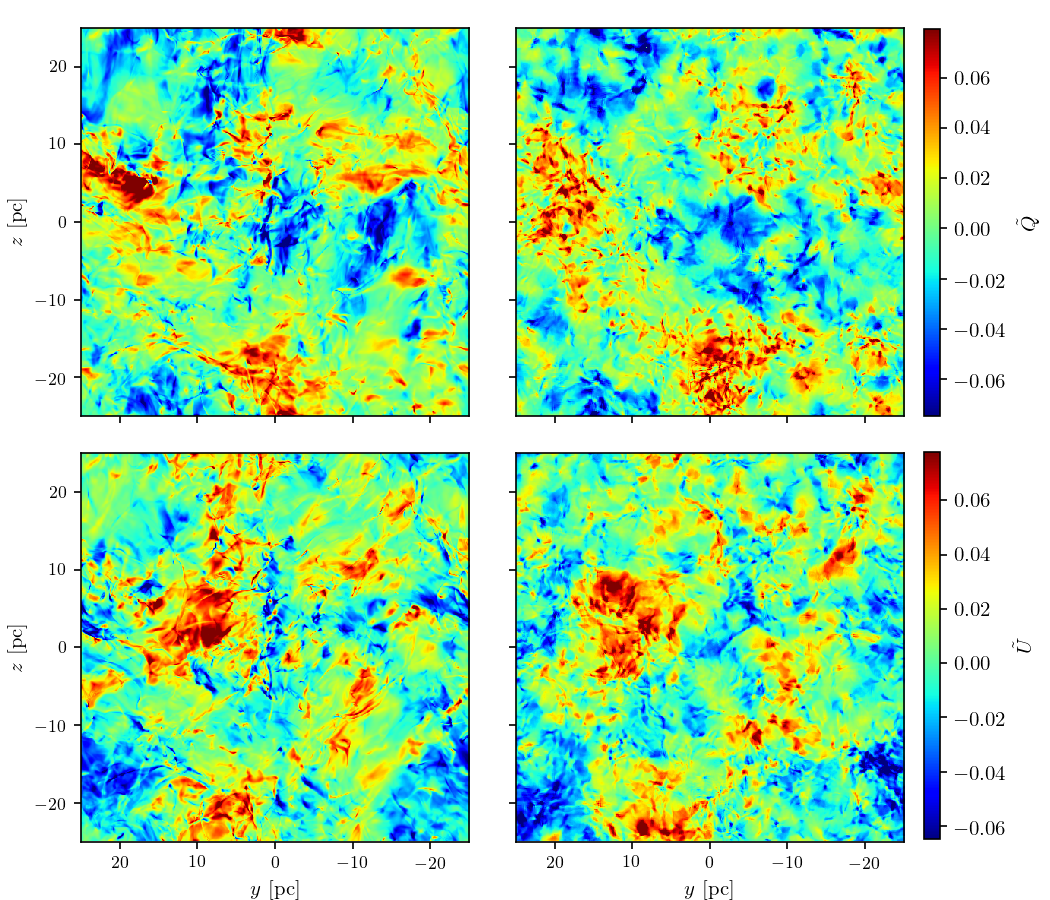}
	\caption{Synthesis of a $\tilde{Q}_{\parallel}+i\tilde{U}_{\parallel}$ map (right column) built from its corresponding RWST description with additional constraints on large-scale components and on a few one-point moments of $\tilde{Q}$ and $\tilde{U}$ maps, to be consistent with those of a reference map (left column). The reference maps shown here are the same as in Fig.~\ref{datasetParallel}.}
	\label{synthesesPlot}
\end{figure*}

If synthetic total intensity maps generated from a RWST description have already been produced in \citet{Allys2019}, in this paper we extend and improve this procedure for complex polarization maps $\tilde{Q} + i\tilde{U}$. We give in this subsection some technical details on the implementation but details on the mathematical formalism can be found in \citet{Bruna2019}.

The generation of synthetic maps is an iterative process that consists in the minimization of a loss function $\mathcal{L}$ in pixel space. Hence this optimization problem is defined in practice in a space of dimension $512^2$. Starting from a realization of a complex Gaussian white noise map $X_0 = \tilde{Q}_0 + i \tilde{U}_0$, successive maps~$\{X_k\}$ are built with a quasi-Newton method. We use a L-BFGS-B algorithm, although we do not impose any boundary to the values of the pixels. Technical details on quasi-Newton methods and L-BFGS-B algorithm can be found in \citet{Fletcher1987} and \citet{Byrd1995}.

In the following, one specific map $X^\mathrm{r} = \tilde{Q}^\mathrm{r} + i\tilde{U}^\mathrm{r}$ of the data set on which the RWST description is built serves as a reference for comparison with the synthetic maps. Because our WST and RWST analyses does not extend to the largest scales, we have no statistical information on these. Thus, we replace the largest scales of $X_0$ with those of $X^\mathrm{r}$ to address this gap in a deterministic way\footnote{We note that such an approach cannot supply the statistical information concerning the couplings between the largest scales with smaller ones.}. Formally, noting $\mathcal{F}[X_0]$ the Fourier transform of $X_0$ initially drawn as a realization of a Gaussian white noise we set:
\begin{equation}
    \mathcal{F}[X_0](\ve{k}) = \mathcal{F}[X^\mathrm{r}](\ve{k}) \text{ for } |\ve{k}| < k_{\mathrm{min}},
\end{equation}
where $k_{\mathrm{min}}$ is the wave number corresponding to the largest scale probed by the WST. For $J = 7$, using Table~\ref{tableScales} we have $k_{\mathrm{min}} \approx 2\pi/171 \approx 0.037$ pixel$^{-1}$. In practice, 25 Fourier modes out of $512^2$ are set by this procedure. Even if in our case this value of $k_{\mathrm{min}}$ is related to an ad hoc modeling decision ($J$ value), we point out that statistical approaches are not always relevant at all scales for the analysis of the diffuse ISM. For example, in a turbulent medium we know that injection scales generally correspond to specific events (e.g., a supernova) for which a statistical description has little meaning, while a statistical description is relevant to describe the inertial range. Moreover, the WST of an image does not adequately characterize its one-point statistics, so in the following we also constrain these for the synthetic maps using $X^\mathrm{r}$ one-point statistics as a reference.

We define the loss function $\mathcal{L}$ of a complex image $X = \tilde{Q}+i\tilde{U}$ as follows:
\begin{align}
    \mathcal{L}[X] = \mathcal{L}_{\mathrm{WST}}[X] + \mu\left(\mathcal{L}_{\mathrm{one-point}}[\tilde{Q}]+\mathcal{L}_{\mathrm{one-point}}[\tilde{U}]\right),
\end{align}
with $\mathcal{L}_{\mathrm{WST}}[X]$ the loss function constraining the WST coefficients of $X$, $\mathcal{L}_{\mathrm{one-point}}[\tilde{Q}]$ (respectively $\mathcal{L}_{\mathrm{one-point}}[\tilde{U}]$) that constraining a few one-point moments of $\tilde{Q}$ (respectively $\tilde{U}$), and $\mu$ a weighting coefficient balancing the importance of one-point moments constraints with that of WST coefficients constraints.
More specifically we set:
\begin{align}
    \mathcal{L}_{\mathrm{WST}}[X] = &\frac{1}{N}\left(\left(S_0[X]-S_0^\mathrm{t}\right)^2 + \sum_{j_1,\theta_1}\left(S_1(j_1,\theta_1)[X] - S_1^\mathrm{t}(j_1,\theta_1)\right)^2
      \right. \nonumber\\ +  & \left.  \sum_{j_1,j_2,\theta_1,\theta_2}\left(S_2(j_1,\theta_2,j_2,\theta_2)[X] - S_2^\mathrm{t}(j_1,\theta_1,j_2,\theta_2)\right)^2\right),
\end{align}
where the "t" superscript refers to the \textit{target} WST coefficients that the synthetic map should have, and $N$ is the total number of WST coefficients. In our case ($J$ = 7 and $\Theta = 8$) we recall that we have $N = 2801$. These target WST coefficients are computed from the RWST coefficients derived from a given data set of ${\tilde{Q}+i\tilde{U}}$ maps (of which $X^{\rm r}$ is part) using the RWST model defined in Eqs.~(\ref{eqRWSTS1}) and (\ref{eqRWSTS2}). We see on this loss function that none of the WST coefficient is privileged and that we do not weigh the differences of WST coefficients by any uncertainty on the target coefficients. This is certainly something that can be improved in future work on WST syntheses.
We finally define:
\begin{align}
    \mathcal{L}_{\mathrm{one-point}}[\tilde{Q}] = \frac1{3}&\left(\left(\frac{\mathcal{M}_2[\tilde{Q}]}{\mathcal{M}_2[\tilde{Q}^\mathrm{r}]} - 1\right)^2 + \left(\mathcal{M}_3[\tilde{Q}] - \mathcal{M}_3[\tilde{Q}^r]\right)^2\right. \nonumber\\ + & \left. \left(\mathcal{M}_4[\tilde{Q}] - \mathcal{M}_4[\tilde{Q}^r]\right)^2\right),
\end{align}
where ${\mathcal{M}_2[\tilde{Q}] = \langle \left(\tilde{Q} - \langle \tilde{Q}\rangle\right)^2 \rangle}$ is an estimation of the variance of the distribution of pixels values in the $\tilde{Q}$ map, ${\mathcal{M}_3[\tilde{Q}] = \langle \left(\tilde{Q} - \langle \tilde{Q}\rangle\right)^3 \rangle / \mathcal{M}_2[\tilde{Q}]^{3/2}}$ is an estimation of its skewness, and ${\mathcal{M}_4[\tilde{Q}] = \langle \left(\tilde{Q} - \langle \tilde{Q}\rangle\right)^4 \rangle / \mathcal{M}_2[\tilde{Q}]^{2}}$ is an estimation of its kurtosis.

\begin{figure*}
	\centering
	\includegraphics[width=\textwidth]{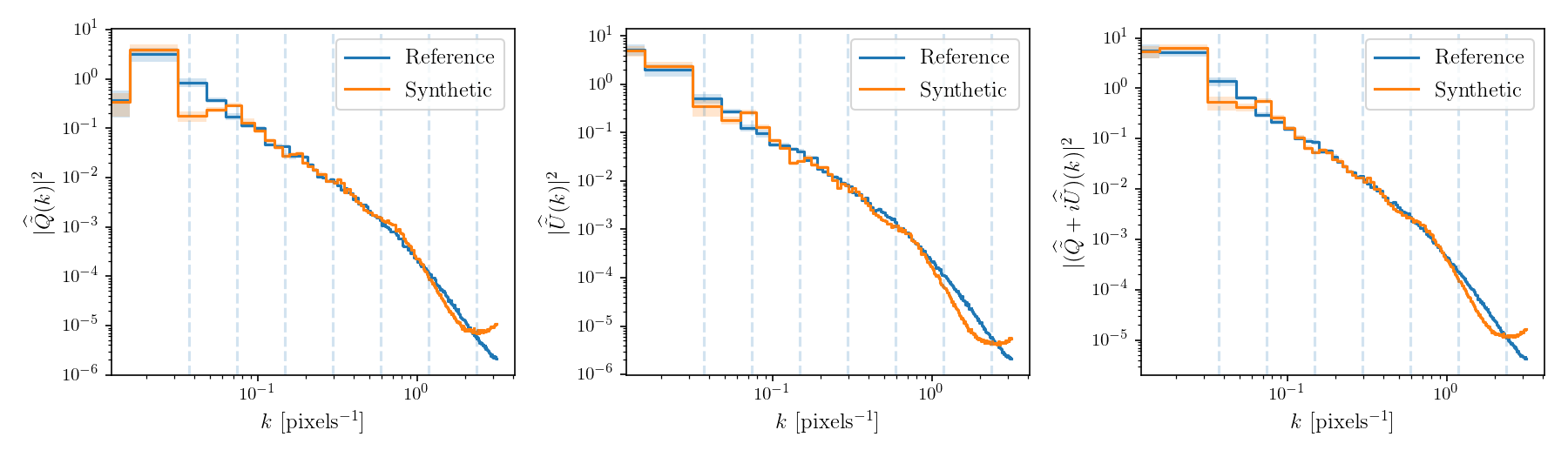}
	\caption{Power spectra of the synthetic $\tilde{Q}_{\parallel}+i\tilde{U}_{\parallel}$ map shown in Fig.~\ref{synthesesPlot}, compared to those of the reference map, for $\tilde{Q}$, $\tilde{U}$, and $|\tilde{Q}+i\tilde{U}|$ (left, middle and right respectively). The vertical dashed lines mark the wavelet central wave numbers corresponding to the scale indices $j=0,\dots,J-1$.}
	\label{synthesesStatsPS}
\end{figure*}

\begin{figure*}
	\centering
	\includegraphics[width=\textwidth]{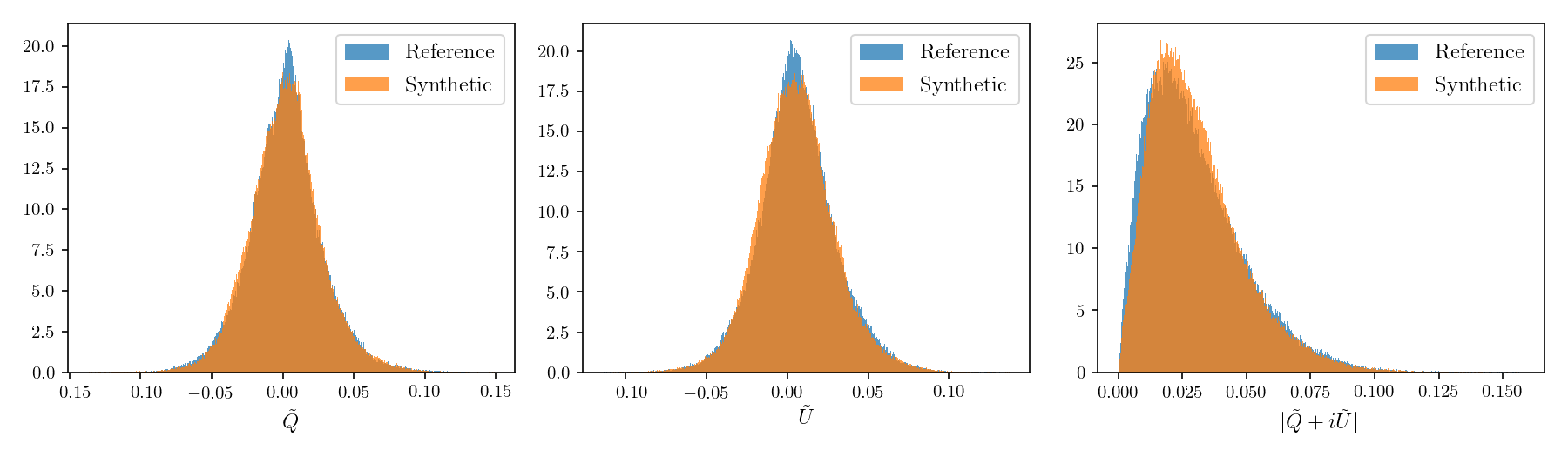}
	\caption{One-point distribution functions of the synthetic $\tilde{Q}_{\parallel}+i\tilde{U}_{\parallel}$ map shown in Fig.~\ref{synthesesPlot}, compared to those of the reference map, for $\tilde{Q}$, $\tilde{U}$, and $|\tilde{Q}+i\tilde{U}|$ (left, middle and right respectively).}
	\label{synthesesStatsPDF}
\end{figure*}

We generate synthetic maps using the $\tilde{Q}_\parallel+i\tilde{U}_\parallel$ RWST description as a target but we also provide equivalent results for the $\tilde{Q}_\bot+i\tilde{U}_\bot$ RWST description in Appendix~\ref{AdditionalFigures}. The optimization algorithm stops when the values of the loss function stagnates between two consecutive iterations. In practice we choose ${\mu = 5\times 10^{-8}}$ and the algorithm stops when ${|\mathcal{L}[X_{k}] - \mathcal{L}[X_{k + 1}]| \lessapprox 10^{-12}}$. These numerical values correspond to a reasonable trade-off between the quality of the syntheses and the execution time of the algorithm. They lead to constrained statistical coefficients with an accuracy better than a few percent on average. Once the optimization is done, we finally filter the modes of the resulting maps at higher wave numbers than the one dimensional Nyquist wave number ($k_N = \pi~\mathrm{pixels}^{-1}$) to avoid unwanted numerical artifacts. Indeed, the WST relies on a bank of Morlet wavelets that does not properly cover this range of frequencies, which results in a total loss function $\mathcal{L}$ that does not properly constrain these modes. Figure~\ref{synthesesPlot} shows a synthetic $\tilde{Q}_\parallel+i\tilde{U}_\parallel$ map (right column) next to its reference map (left column). The overall appearance of the synthetic maps is satisfactory. The largest scales are roughly consistent with those of the reference maps\footnote{Since the WST constraints leave the largest scales of $X_0$ unchanged, this shows that the impact of the one-point moments constraints on these scales is minor.}, but we also see at intermediate and small scales, which are the truly synthetic scales, consistent filamentary patterns and dynamic ranges.

\subsection{One-point and two-point statistics of synthetic maps}

By construction, the synthetic maps we built have the same WST statistics as the one prescribed by the RWST description of the $\tilde{Q}_\parallel+i\tilde{U}_\parallel$ data set, similar large scales and some identical one-point moments compared to the reference maps. We may wonder if elementary one-point and two-point statistics are fully consistent with the ones of the reference maps.

Figure~\ref{synthesesStatsPS} shows the power spectra of $\tilde{Q}$, $\tilde{U}$, and $\tilde{Q}+i\tilde{U}$ up to the one dimensional Nyquist wave number $k_N$ for both the reference maps and the syntheses shown in Fig.~\ref{synthesesPlot}. The power spectra are computed by binning the squared amplitudes of Fourier modes with respect to the modulus of the corresponding wave number $k$. We use a regular binning in $k$ and the estimations of the power spectra are computed as the means for each bin. We also represent standard deviations of the mean per bin. We see that the power spectra as well as their standard deviations are in good agreement for the three variables $\tilde{Q}$, $\tilde{U}$, and $\tilde{Q}+i\tilde{U}$ for all scales except the smallest ones. These discrepancies at small scales take the form of a lack followed by an excess of power in the syntheses for wave numbers approaching $k_N$. We interpret this as the result of poorly constrained modes close to $k_N$ in the optimization process. It is not surprising to reproduce the power spectrum of $\tilde{Q}+i\tilde{U}$ maps as the $\bar{S}_1$ and $\bar{S}_2$ coefficients constrain the power spectrum of $\tilde{Q}+i\tilde{U}$ (see discussion in Sect.~\ref{wstSubsection}). However we point out that we correctly reproduce the power spectra of $\tilde{Q}$ and $\tilde{U}$ taken separately. Still we note that we did not investigate cases where $\tilde{Q}$ and $\tilde{U}$ have very different power spectra.

In Fig.~\ref{synthesesStatsPDF}, we compare the one-point distribution functions of the reference maps and the synthetic maps for $\tilde{Q}$, $\tilde{U}$, and $|\tilde{Q}+i\tilde{U}|$. These are also in fairly good agreement for all the variables. We notably successfully reproduce the tails of these distributions, this must be due to the combined constraints on WST coefficients and one-point moments. One could surely enhance the agreement between the reference and synthetic maps by taking into account higher order moments in the $\mathcal{L}_{\mathrm{one-point}}$ loss functions.

We point out that such syntheses were generated using RWST descriptions that comprise 127 coefficients (see Sect.~\ref{RWSTsubsection}). Adding to that the largest scales that were set in a deterministic way as well as the constraints on the one-point moments of $\tilde{Q}$ and $\tilde{U}$ maps, we end up with a total of 158 coefficients to generate $512 \times 512$ complex $\tilde{Q}+i\tilde{U}$ maps that are in good visual agreement with the maps of the original data set and successfully reproduce their one-point and two-point statistics. Excluding the largest scales, these sets of statistical coefficients not only describe the statistical properties of these maps, but they also define a statistical model of the mechanism that generated these data. In a mathematical wording, these coefficients model the probability distribution of the random field of which these maps are realizations. Although these statistical syntheses will definitely benefit from more dedicated work, they already are a promising avenue to address the construction of the statistical model of polarization maps that we seek.

%%%%%%%%%%%%%%%%%%%%%%%%%%%%%%%%%%%%%%%%%%%%%%%%%%%%
%%%%%%%%%%%%%%%%%%%%%%%%%%%%%%%%%%%%%%%%%%%%%%%%%%%%
%% SECTION ?: CONCLUSION
%%%%%%%%%%%%%%%%%%%%%%%%%%%%%%%%%%%%%%%%%%%%%%%%%%%%
%%%%%%%%%%%%%%%%%%%%%%%%%%%%%%%%%%%%%%%%%%%%%%%%%%%%

\section{Conclusions and perspectives}
\label{SectionConclusions}

In this paper, we extended the WST analysis to maps of polarized thermal emission from interstellar dust, using $512\times512$~pixels Stokes $I$, $Q$, and $U$ maps built from a numerical simulation of MHD turbulence designed to reproduce typical properties of the diffuse ISM. To alleviate the fact that Stokes $Q$ and $U$ rely on the definition of an arbitrary reference frame, and to remove the zeroth-order impact of the matter distribution on their properties and thus focus on the statistics of the magnetic field, the WST was applied to complex Stokes maps $\tilde{Q}+i\tilde{U}$ that are normalized by $I+P$. To study the contributions of the polarization fraction $p$ and of the polarization angle $\psi$ to the statistical properties of these complex Stokes maps, we also applied the WST to the corresponding maps of $p$ and $\exp(2i\psi)$. We finally analyzed "Gaussianized" complex Stokes maps obtained after phase randomization.

The WST gives a low-variance statistical description of these complex and real maps through typically a few thousand coefficients indexed in terms of orientations and scales. These coefficients capture the power spectra of the maps and characterize couplings between oriented scales. WST coefficients for maps of $\tilde{Q}+i\tilde{U}$, $p$, and $\exp(2i\psi)$ present a striking regularity when taken as functions of the sole angular variables. This is very much in line with what we observed in \citet{Allys2019} for column density and total intensity maps, and in fact the same functional form introduced in that paper can be used to fit the angular dependencies of the WST coefficients of polarization maps studied here, thus extending the RWST model introduced in \citet{Allys2019}. The RWST yields a statistical description of polarization maps that quantifies their multiscale properties in terms of isotropic and anisotropic contributions, all the while requiring more than one order of magnitude fewer coefficients than the WST.

In the rest of this section, we summarize the main results of our work, then highlight a few perspectives. The RWST analysis allowed us to identify statistical characteristics that exhibit the dependence of the map structure on the orientation of the mean magnetic field and quantify the non-Gaussianity of data.
\begin{itemize}
   \item The overall level of first order coefficients depends on the orientation of the mean magnetic field with respect to the line of sight. For $\tilde{Q}+i\tilde{U}$ maps, $\hat{S}_1^{\mathrm{iso}} + \log_2(\bar{S}_0)$ coefficients are larger when the mean magnetic field is in the plane of the sky, while for $\exp(2i\psi)$ maps $\hat{S}_1^{\mathrm{iso}}$ is larger when the mean magnetic field is along the line of sight.\\
    \item $\hat{S}_1^{\mathrm{aniso}}$ coefficients quantify the statistical anisotropy of the maps. When the mean magnetic field is parallel to the line of sight $\hat{S}_1^{\mathrm{aniso}}$ coefficients are negligible, while when the mean magnetic field is in the plane of the sky they allow us to identify the direction of anisotropy at each scale. For the MHD simulation we analyzed, this direction is orthogonal to the direction of the mean magnetic field for both the $\tilde{Q}_\bot+i\tilde{U}_\bot$ and $\exp(2i\psi)_\bot$ maps.\\
    \item Second order RWST coefficients clearly exhibit the non-Gaussianity of $\tilde{Q}+i\tilde{U}$ maps (although this is also visible to a lesser extent in first order coefficients). While the randomized $R[\tilde{Q}+i\tilde{U}]$ data sets present characteristic properties of scale invariant Gaussian fields (invariance of $\hat{S}_2^{\mathrm{iso},1}$ as a function of the scale difference $j_2-j_1$ and a quick decrease to zero of $\hat{S}_2^{\mathrm{iso},2}$ as this scale difference increases), the $\hat{S}_2^{\mathrm{iso},1}$ and $\hat{S}_2^{\mathrm{iso},2}$ coefficients for the corresponding $\tilde{Q}+i\tilde{U}$ data sets show clearly different patterns. In particular, the strictly positive values of $\hat{S}_2^{\mathrm{iso},2}$ at large $j_2-j_1$ are interpreted as signatures of the filamentary structure of the maps.
\end{itemize}

We have used the RWST approach to synthesize $\tilde{Q}+i\tilde{U}$ maps. Combining the RWST coefficients with additional constraints, we obtained synthetic maps that statistically match the original maps. The agreement demonstrates the comprehensiveness of the statistical description provided by the RWST. The additional constraints include large scale modes that cannot be described statistically with the limited amount of samples we have worked with, as well as statistical constraints on the one-point distribution function.

In this paper, to establish the methodology, we have worked with noise-free maps computed from numerical simulations of MHD turbulence. A future extension would be to apply it to observations of polarized thermal emission from dust. To do this, we need to learn how to handle data noise. Indeed, while signal-to-noise ratios in total intensity for both {\it Herschel} and {\it Planck} maps are quite high, this is not the case for currently available polarization data. Studying how data noise affects the WST coefficients would demand to repeat the analysis of MHD simulations with noise added to the dust signal. Once this difficulty is overcome, we may use the RWST to define a metric to compare observations with simulations and phenomenological models. This will be a stepping stone towards a more refined physical interpretation of the RWST coefficients. A main motivation would be to use the RWST to characterize magnetized interstellar turbulence.

Throughout this work we chose to work with Stokes $I$, $Q$, and $U$ maps to analyze the polarization of dust thermal emission as astronomers do for Galactic astrophysics. In the framework of CMB data analysis, polarization is more usually characterized through E and B modes. Thus, it would be interesting to apply the RWST to E and B maps. This will lead to  a physically motivated statistical model that would include observational constraints such as the E/B power asymmetry and the correlation between E-modes of the polarization and total intensity~\citep{planck2016-l11A}. Once this is achieved, the RWST may be used to synthesize dust polarization maps that would help to assess and optimize component separation methods for CMB data analysis.

\begin{acknowledgements}
We thank the anonymous referee for a very helpful report. We gratefully acknowledge V. Jelić for pointing out to us the complex-valued description of polarization maps. We also acknowledge fruitful discussions with T. Marchand, J.-F. Cardoso, A. Frolov, S. Zhang, and S. Mallat that enriched this work in various ways. This research was supported by the Agence Nationale de la Recherche (project BxB: ANR-17-CE31-0022). F. Levrier and F. Boulanger acknowledge support from the European Research Council under the European Union’s Horizon 2020 Research \& Innovation Framework Programme / ERC grant agreement ERC-2016-ADG-742719. B. Regaldo-Saint Blancard acknowledges support from the Centre National d'Etudes Spatiales (CNES).
\end{acknowledgements}

%
% Begin Appendix
%
\begin{appendix}

%%%%%%%%%%%%%%%%%%%%%%%%%%%%%%%%%%%%%%%%%%%%%%%%%%%%
%%%%%%%%%%%%%%%%%%%%%%%%%%%%%%%%%%%%%%%%%%%%%%%%%%%%
%% APPENDIX 1: ?
%%%%%%%%%%%%%%%%%%%%%%%%%%%%%%%%%%%%%%%%%%%%%%%%%%%%
%%%%%%%%%%%%%%%%%%%%%%%%%%%%%%%%%%%%%%%%%%%%%%%%%%%%

\section{The RWST model in terms of Fourier series expansions}
\label{rwstGenerality}

The RWST model can be rephrased in terms of truncated Fourier series. Let us consider the $\log_2(\bar{S}_1)$ coefficients at a given scale $j_1$ and write $f_{j_1}(\theta_1)$ the corresponding angular model\footnote{The same kind of reasoning holds for $\log_2(\bar{S}_2)$ coefficients.}. If one assumes that there are no more than one privileged direction $\theta^{\mathrm{ref}, 1}(j_1)$ in the maps we are dealing with, we can write generally  $f_{j_1}(\theta_1)$ as a Fourier series expansion using the natural $2\pi$-periodicity of this function:
\begin{align}
     f_{j_1}(\theta_1)= a_0(j_1) + \sum_{k = 1}^{+\infty} & \left[a_k(j_1)\cos\left(k\left[\theta_1 - \theta^{\mathrm{ref}, 1}\left(j_1\right)\right]\right) + \nonumber \right. \\  & \left. b_k(j_1)\sin\left(k\left[\theta_1 - \theta^{\mathrm{ref}, 1}\left(j_1\right)\right]\right)\right].
\end{align}
Assuming a mirror symmetry with respect to the potential reference direction, we expect $\{b_k\}$ coefficients to vanish (which is the case in practice for our data). Adding to that the {$\pi$-periodicity} identified in Sect.~\ref{regularitySubsection} the angular model reduces to:
\begin{equation}
    f_{j_1}(\theta_1) = a_0(j_1) + \sum_{k = 2}^{+\infty}a_k(j_1)\cos\left(k\left[\theta_1 - \theta^{\mathrm{ref}, 1}\left(j_1\right)\right]\right).
\end{equation}
Finally, the smoothness of the patterns presented in Sect.~\ref{regularitySubsection} implies a fast decrease of the amplitudes of the harmonics. Truncating the expansion after the second term and writing ${a_0(j_1) = \hat{S}_1^{\mathrm{iso}}(j_1)}$ and ${a_2(j_1) = \hat{S}_1^{\mathrm{aniso}}(j_1)}$ we end up with the RWST model of Eq.~(\ref{eqRWSTS1}).

%%%%%%%%%%%%%%%%%%%%%%%%%%%%%%%%%%%%%%%%%%%%%%%%%%%%
%%%%%%%%%%%%%%%%%%%%%%%%%%%%%%%%%%%%%%%%%%%%%%%%%%%%
%% APPENDIX 2: ?
%%%%%%%%%%%%%%%%%%%%%%%%%%%%%%%%%%%%%%%%%%%%%%%%%%%%
%%%%%%%%%%%%%%%%%%%%%%%%%%%%%%%%%%%%%%%%%%%%%%%%%%%%

\section{Phase randomization}
\label{PhaseRandomization}

We give some technical details concerning the phase randomization process of a discretized scalar field.

\begin{figure}
	\centering
	\includegraphics[width=0.5\textwidth]{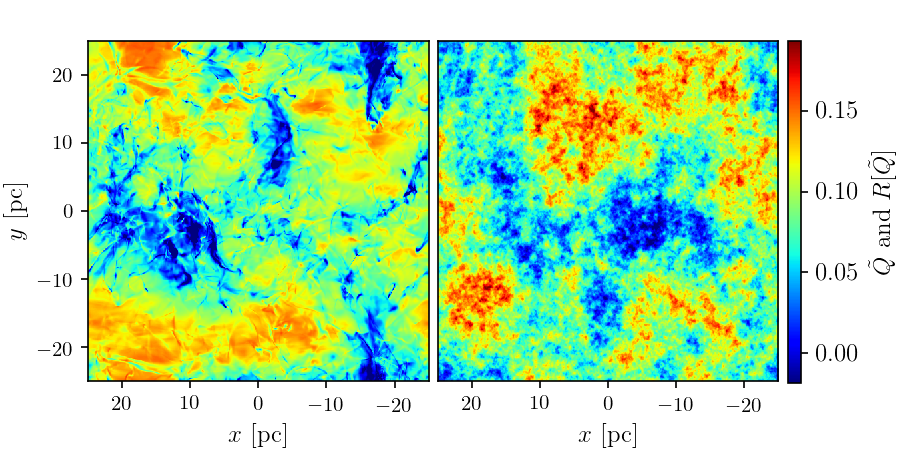}
	\caption{Example of a phase randomized $R[\tilde{Q}_\bot]$ map (right) next to its corresponding original $\tilde{Q}_\bot$ map (left).}
	\label{datasetRandomized}
\end{figure}

Let us consider a 2D discretized real scalar field $f : \Omega \rightarrow \mathbb{R}$ with $\Omega =\llbracket 0, N - 1 \rrbracket^2$. Its discrete Fourier transform reads for a given mode $\ve{k}=(k_x,k_y)\in \Omega$:
\begin{equation}
	\hat{f}(\ve{k}) = \sum_{\ve{r}\in\Omega}{f(\ve{r})\exp\left(-2\pi i \frac{\ve{r}.\ve{k}}{N}\right)},
\end{equation}
where $\ve{r}\cdot\ve{k}=xk_x+yk_y$. We also recall the inverse discrete Fourier transform relation for the chosen convention:
\begin{equation}
	f(\ve{r}) = \frac1{N^2}\sum_{\ve{k}\in\Omega}{\hat{f}(\ve{k})\exp\left(2\pi i \frac{\ve{r}.\ve{k}}{N}\right)}.
\end{equation}
The phase randomization of $f$ consists in defining a new field $g$ in Fourier space such that:
\begin{equation}
	\hat{g}(\ve{k}) = |\hat{f}(\ve{k})|e^{i\phi(\ve{k})},
\end{equation}
for every mode $\ve{k}\in\Omega$ and where $\phi$ is a realization of a uniform random phase such as defined in \citet{Galerne2011}, that is, defining $\Omega_0 = \{(0,N/2),(N/2,0),(N/2,N/2)\}$ if $N$ is even, and $\Omega_0=\varnothing$ otherwise, $\phi$ verifies:
\begin{enumerate}
	\item $\forall \ve{k}\in \Omega \setminus \Omega_0, \phi(-\ve{k}) = -\phi(\ve{k})$ (we extend the domain of $\phi$ to $\mathbb{Z}^2$ using periodic boundary conditions).
	\item $\forall$ $\ve{k}\in\Omega$, $\phi(\ve{k})$ is drawn uniformly and independently in $[0,2\pi)$ (the independence holds to the extent of the first relation).
	\item $\forall \ve{k} \in \Omega_0, \phi(\ve{k})$ is drawn uniformly and independently in the set $\{0,\pi\}$.
\end{enumerate}
These relations ensure that $g$ will be a real field, and that the mean of $g$ will be equal to the absolute value of the mean of $f$ (since condition 1. gives $\phi(0)=0$). Such a process obviously defines a field $g$ which preserves the power spectrum of $f$. For reference, we show an example of a phase randomized  $R[\tilde{Q}_\bot]$ map in Fig.~\ref{datasetRandomized}.

%%%%%%%%%%%%%%%%%%%%%%%%%%%%%%%%%%%%%%%%%%%%%%%%%%%%
%%%%%%%%%%%%%%%%%%%%%%%%%%%%%%%%%%%%%%%%%%%%%%%%%%%%
%% APPENDIX 3: ?
%%%%%%%%%%%%%%%%%%%%%%%%%%%%%%%%%%%%%%%%%%%%%%%%%%%%
%%%%%%%%%%%%%%%%%%%%%%%%%%%%%%%%%%%%%%%%%%%%%%%%%%%%

\section{Additional results}
\label{AdditionalFigures}

\subsection{Additional terms in the RWST model}
\label{LatticeCorrections}

Following the appendix C of \citet{Allys2019}, we enhance the RWST model defined in Eq.~(\ref{eqRWSTS1}) and (\ref{eqRWSTS2}) by adding so-called \textit{lattice} terms related to pixellization effects at small scales for first order coefficients and a supplementary harmonic of the angular modulation of the second order WST coefficients. This enhanced RWST model of the angular dependency of the WST coefficients becomes, for $\bar{S}_1$ coefficients:
\begin{align}
    \label{eqRWSTS1Lat}
    \log_2\left[\bar{S}_1\left(j_1, \theta_1\right)\right] = &\hat{S}_1^{\mathrm{iso}}(j_1) \nonumber \\ + &\hat{S}_1^{\mathrm{aniso}}(j_1)\cos\left(2\left[\theta_1 - \theta^{\mathrm{ref}, 1}\left(j_1\right)\right]\right)\nonumber \\
    + &\hat{S}_1^{\mathrm{lat}, 1}(j_1)\cos\left(4\theta_1\right) + \hat{S}_1^{\mathrm{lat}, 2}(j_1)\cos\left(8\theta_1\right),
\end{align}
where the additional lattice terms $\hat{S}_1^{\mathrm{lat}, 1}$ and $\hat{S}_1^{\mathrm{lat}, 2}$ quantify angular modulations that are respectively $\pi/2$ and $\pi/4$-periodic and aligned with the main directions of the lattice. Similarly the enhanced RWST model of $\bar{S}_2$ coefficients is:
\begin{align}
    \label{eqRWSTS2Lat}
       \log_2  & \left[ \bar{S}_2 \left(j_1, \theta_1, j_2, \theta_2\right)\right] =\hat{S}_2^{\mathrm{iso}, 1}(j_1, j_2) \nonumber \\ + & \hat{S}_2^{\mathrm{iso}, 2}(j_1, j_2)\cos\left(2\left[\theta_1 - \theta_2\right]\right)\nonumber \\
       + & \hat{S}_2^{\mathrm{iso}, 3}(j_1, j_2)\cos\left(4\left[\theta_1 - \theta_2\right]\right)\nonumber \\
       + & \hat{S}_2^{\mathrm{aniso},1}(j_1, j_2)\cos\left(2\left[\theta_1 - \theta^{\mathrm{ref}, 2}\left(j_1, j_2\right)\right]\right)\nonumber \\ + & \hat{S}_2^{\mathrm{aniso},2}(j_1, j_2)\cos\left(2\left[\theta_2- \theta^{\mathrm{ref}, 2}\left(j_1, j_2\right)\right]\right),
\end{align}
where the additional term $\hat{S}_2^{\mathrm{iso}, 3}$ measures the amplitude of an additional harmonic of the $\theta_1-\theta_2$ angular modulation that is $\pi/2$-periodic.

\begin{figure}
	\centering
	\includegraphics[width=0.9\hsize]{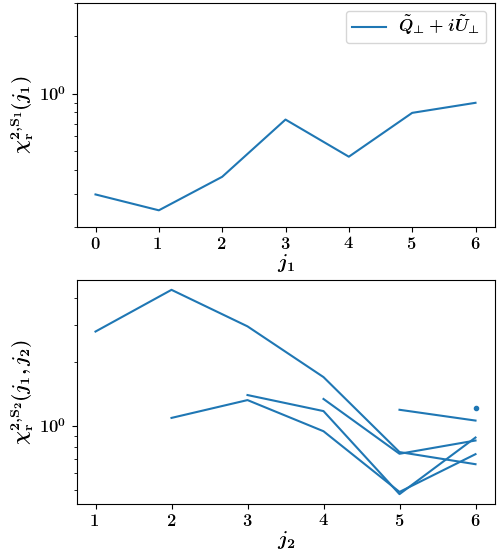}
	\caption{Reduced chi square $\chi_\mathrm{r}^{2, \mathrm{S_1}}(j_1)$ (top) and $\chi_\mathrm{r}^{2, \mathrm{S_2}}(j_1, j_2)$ (bottom) associated with the RWST fits of the WST coefficients that take into account lattice terms (see Eqs.~(\ref{eqRWSTS1Lat}) and (\ref{eqRWSTS2Lat})) for the $\tilde{Q}_\bot+i\tilde{U}_\bot$ data set. Each curve in the $\chi_\mathrm{r}^{2, \mathrm{S_2}}(j_1, j_2)$ plot corresponds to a fixed $j_1$ value while $j_2$ ranges from $j_1 + 1$ to $J-1=6$. For $j_1 = J - 2$, the curve is reduced to a single dot on the figure. We use logarithmic scales for better visibility.}
	\label{rwstChi2rLat}
\end{figure}

\begin{figure}
	\centering
	\includegraphics[width=\hsize]{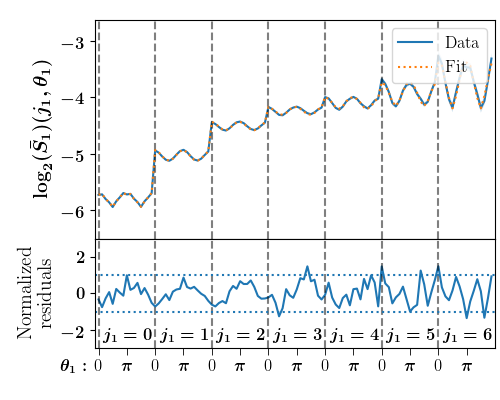}
	\caption{Same as Fig.~\ref{wstS1} but for the RWST fit corresponding to the model of Eq.~(\ref{eqRWSTS1Lat}) including lattice terms.}
	\label{wstS1Lat}
\end{figure}

\begin{figure*}
	\centering
	\includegraphics[width=\textwidth]{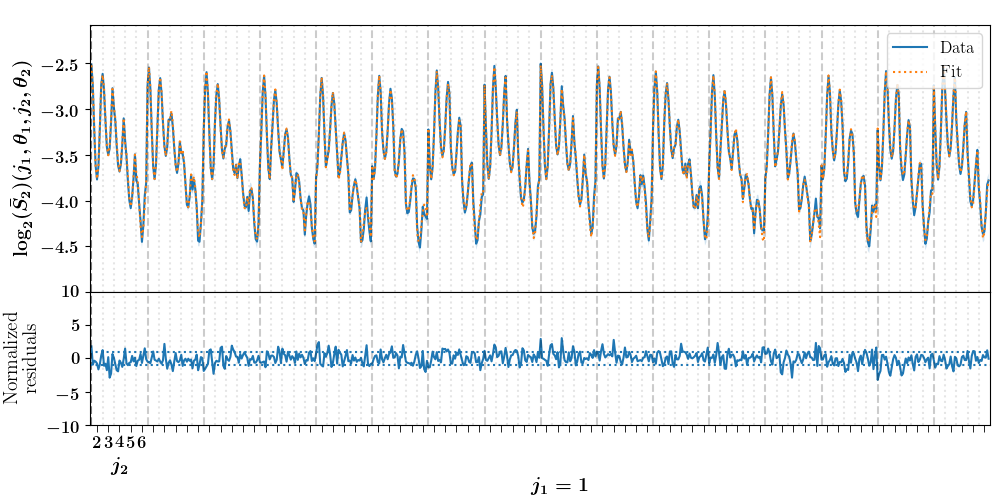}
	\caption{Same as Fig.~\ref{wstS2} but for the RWST fit corresponding to the model of Eq.~(\ref{eqRWSTS1Lat}) including lattice terms.}
	\label{wstS2Lat}
\end{figure*}

These additional terms do not affect the values of the RWST coefficients discussed in the main body of the paper and significantly improve reduced chi square $\chi_\mathrm{r}^{2, \mathrm{S_1}}$ and $\chi_\mathrm{r}^{2, \mathrm{S_2}}$ at small scales as shown in Fig.~\ref{rwstChi2rLat}. These chi square functions are to be compared to those of Fig.~\ref{rwstChi2r}. Figures~\ref{wstS1Lat} and \ref{wstS2Lat} show greatly improved normalized residuals compared to the previous ones shown in Figs.~\ref{wstS1} and \ref{wstS2}. In particular at $j_1 = 0$ we no longer observe the strong angular pattern seen in Fig.~\ref{wstS1}.

\begin{figure*}
	\centering
	\includegraphics[width=\textwidth]{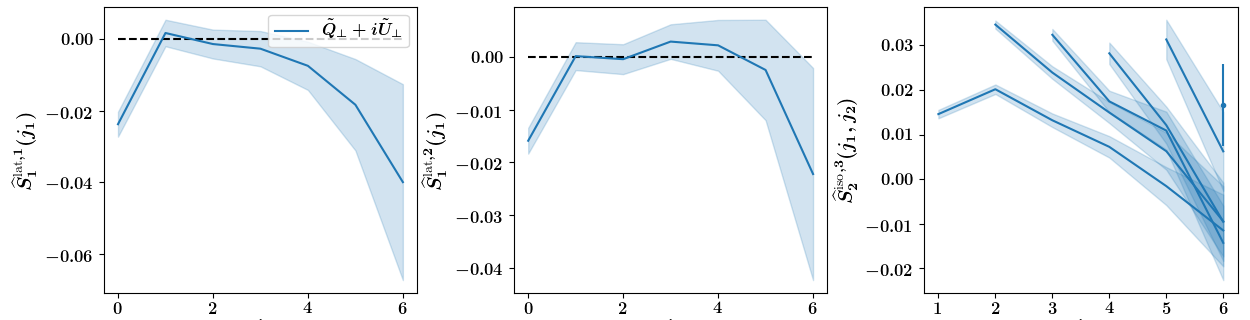}
	\caption{$\hat{S}_1^{\mathrm{lat}, 1}(j_1)$ (left column), $\hat{S}_1^{\mathrm{lat}, 2}(j_1)$ (middle column) and $\hat{S}_2^{\mathrm{iso}, 3}(j_1,j_2)$ (right column) RWST coefficients for the $\tilde{Q}_\bot+i\tilde{U}_\bot$ data set. In the $\hat{S}_2^{\mathrm{iso}, 3}$ plot, each curve corresponds to a fixed $j_1$ value while $j_2$ ranges from $j_1 + 1$ to $J-1=6$. For $j_1 = J - 2$, the curve is reduced to a single dot on the figure.}
	\label{RWSTLatCoeffs}
\end{figure*}

For completeness, we show in Fig.\ref{RWSTLatCoeffs} the additional RWST terms given for the example of the $\tilde{Q}_\bot+i\tilde{U}_\bot$ data set.

\subsection{Syntheses for $\tilde{Q}_{\bot}+i\tilde{U}_{\bot}$}

\begin{figure*}
	\centering
	\includegraphics[width=0.6\textwidth]{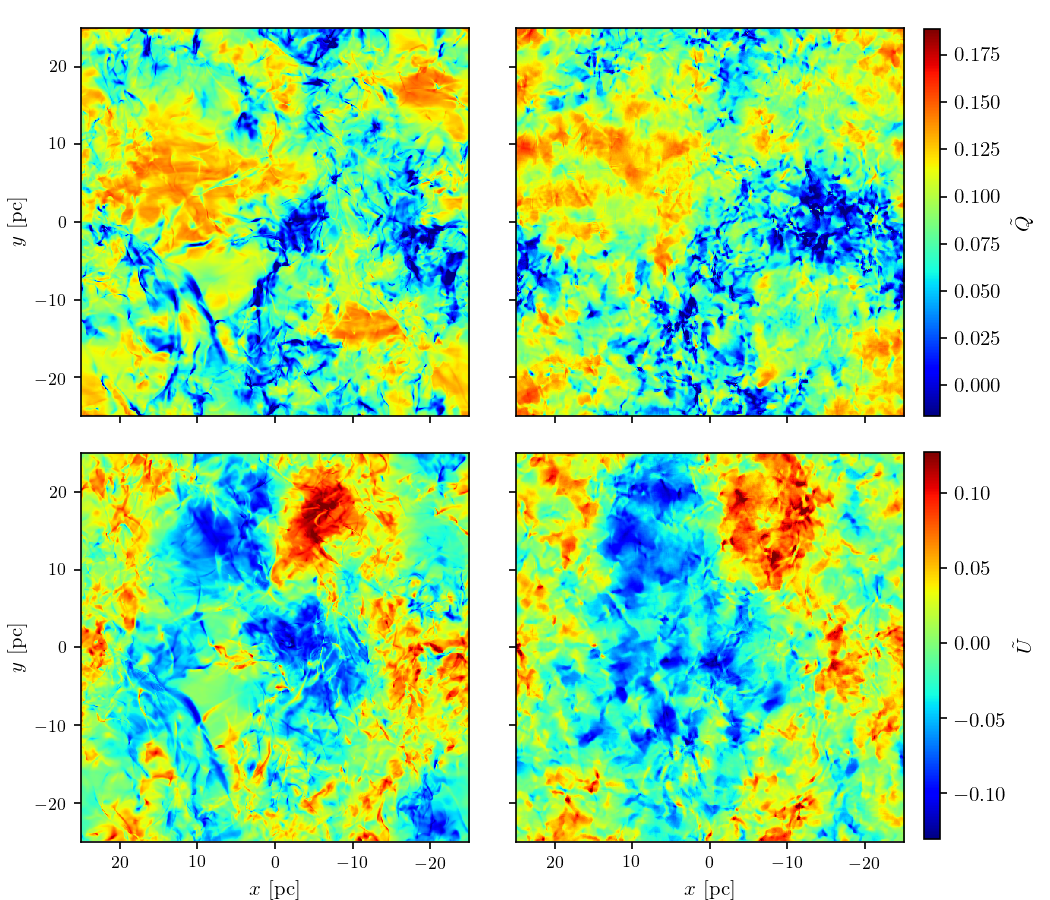}
	\caption{Same as Fig.~\ref{synthesesPlot} but for the $\bot$ case. The reference maps shown here are the same maps as in Fig.~\ref{datasetOrtho}.}
	\label{synthesesPlotOrtho}
\end{figure*}

\begin{figure*}
	\centering
	\includegraphics[width=\textwidth]{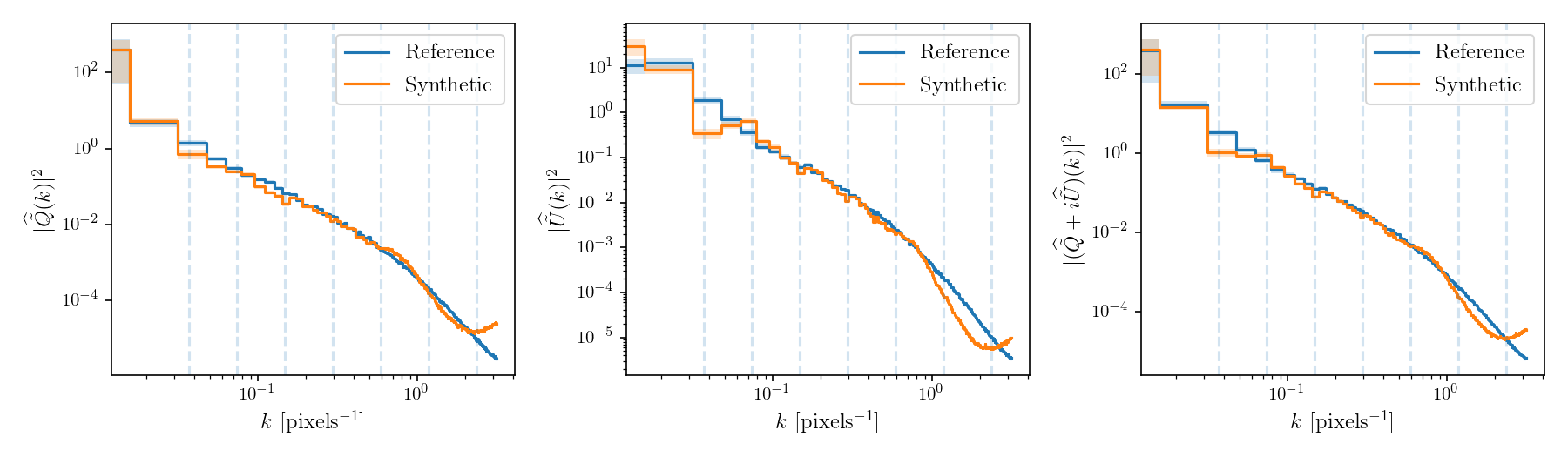}
	\caption{Same as Fig.~\ref{synthesesStatsPS} but for the $\bot$ case.}
	\label{synthesesStatsPSOrtho}
\end{figure*}

\begin{figure*}
	\centering
	\includegraphics[width=\textwidth]{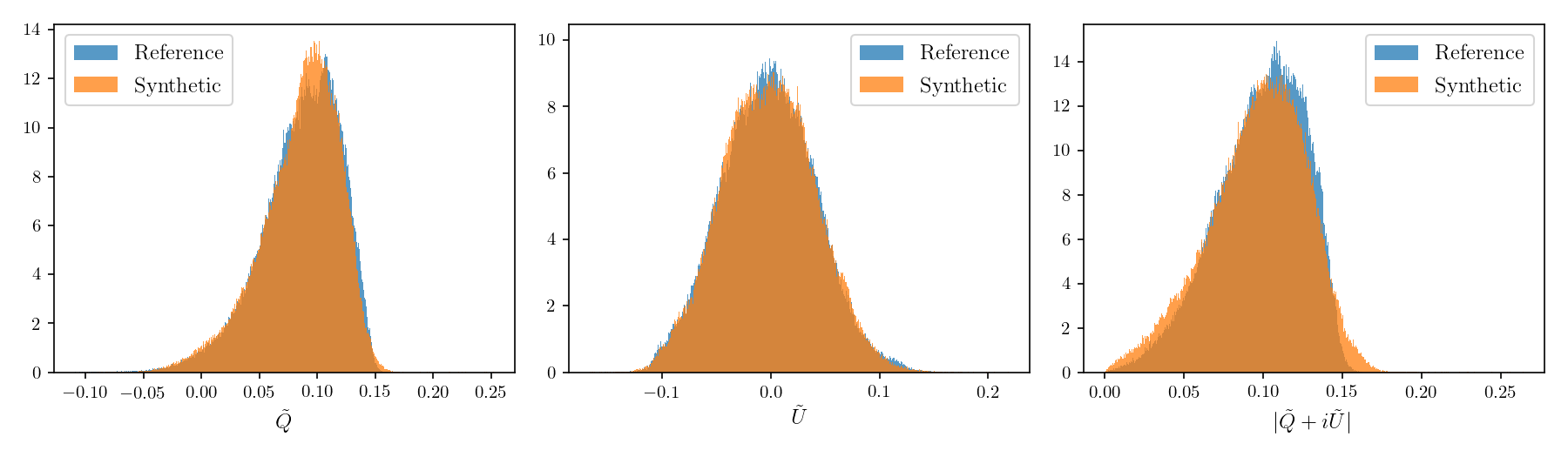}
	\caption{Same as Fig.~\ref{synthesesStatsPDF} but for the $\bot$ case.}
	\label{synthesesStatsPDFOrtho}
\end{figure*}

We show in Figs.~\ref{synthesesPlotOrtho}, \ref{synthesesStatsPSOrtho}, and \ref{synthesesStatsPDFOrtho} the maps, power spectra, and distribution functions of synthetic maps in the $\bot$ case that are analogous to those of Figs.~\ref{synthesesPlot}, \ref{synthesesStatsPS}, and \ref{synthesesStatsPDF}.

\end{appendix}

% For the bibliography, at the end
\bibliographystyle{Bibtex/aa} % style aa.bst
\bibliography{Bibtex/Bibliography.bib}

\end{document}